\documentclass[AMA,STIX1COL]{WileyNJD-v2}
\newcommand\numberthis{\addtocounter{equation}{1}\tag{\theequation}}

\usepackage{tablefootnote}
\usepackage{adjustbox}
\usepackage{xcolor}
\usepackage{ulem}
\colorlet{RED}{red}
\colorlet{GRAY}{gray}

\articletype{Research Article}%

\received{}
\revised{}
\accepted{}
        
\begin{document}

\title{A Bayesian hierarchical meta-analytic method for modelling surrogate relationships that vary across treatment classes using aggregate data}

\author[1]{Tasos Papanikos}
\author[2]{John Thompson}
\author[1]{Keith Abrams}
\author[3]{Nicolas St{\"a}dler}
\author[4,5]{Oriana Ciani}
\author[4]{\newline Rod Taylor}
\author[1]{Sylwia Bujkiewicz}

\authormark{Papanikos \textsc{et al}}

\address[1]{Biostatistics Group, Department of Health Sciences, University of Leicester, Leicester, UK}
\address[2]{Genetic Epidemiology Group, Department of Health Sciences, University of Leicester, Leicester, UK}
\address[3]{Roche Innovation Centre, F. Hoffmann-La Roche Ltd, Basel, Switzerland}
\address[4]{Institute of Heath Research, University of Exeter Medical School, Exeter, UK}
\address[5]{Centre for Research on Health and Social Care Management, Bocconi University, Milan, Italy}

\corres{Tasos Papanikos,  \email{ap659@leicester.ac.uk}}

\presentaddress{Biostatistics Research Group \\ Department of Health Sciences \\ University of Leicester \\ George Davies Centre \\ University Road \\ Leicester \\ LE1 7RH \\ United Kingdom}

\abstract[Abstract]{Surrogate endpoints play an important role in drug development when they can be used to measure treatment effect
	early compared to the final clinical outcome and to predict clinical benefit or harm. Such endpoints are assessed for their predictive value of clinical
	benefit by investigating the surrogate relationship between treatment effects on the surrogate and final outcomes using
	meta-analytic methods. When surrogate relationships vary across treatment classes, such validation may fail due to
	limited data within each treatment class. In this paper, two alternative Bayesian meta-analytic methods are introduced which allow for borrowing of
	information from other treatment classes when exploring the surrogacy in a particular class. The first approach extends
	a standard model for the evaluation of surrogate endpoints to a hierarchical meta-analysis model assuming full exchangeability of surrogate relationships across
	all the treatment classes, thus facilitating borrowing of information across the classes. The second method is able to relax
	this assumption by allowing for partial exchangeability of surrogate relationships across treatment classes to avoid
	excessive borrowing of information from distinctly different classes. We carried out a simulation study to assess
	the proposed methods in nine data scenarios and compared them with subgroup analysis using the standard model within each treatment class. We also applied the methods to an
	illustrative example in colorectal cancer which led to obtaining the parameters describing the surrogate relationships with
	higher precision.
}

\keywords{surrogate endpoints, meta-analysis, hierarchical models, partial exchangeability, treatment classes}

\maketitle

	\section{Introduction}\label{Introduction}
	New advances in science have led to discovering of promising therapies which often are targeted to specific patient populations, for example defined by a genetic biomarker. This leads to clinical trials of smaller size, whilst the increased effectiveness of these therapies reduces the number of events or deaths and consequently lead to measurement of treatment effect on overall survival (OS) with large uncertainty. Therefore, surrogate endpoints allowing the measurement of treatment effect with higher precision have been investigated to accelerate the availability of these treatments to the patients. 
	These  alternative endpoints often can be considered a cost effective replacement of final clinical outcome, as they are particularly useful when they can be measured earlier, easier, more frequently compared to the final clinical endpoint or if they require smaller sample size and shorter follow up times \cite{burzykowski2006evaluation}.
	
	Potential surrogate endpoints have been investigated as candidate endpoints in clinical trials in a number of disease areas. However, before these candidate endpoints are used, either as primary endpoints in trial design or in regulatory decision making, they need to be validated \cite{fleming2012biomarkers}. In practice, the most common approach to validate a candidate outcome is to examine whether it satisfies three levels of association, proposed by the International Conference on Harmonisation Guidelines on Statistical Principles for Clinical Trials (ICH) \cite{phillips2003e9}. First, the biological plausibility of the association of the surrogate and final outcomes is investigated which, involves biological rather than statistical considerations. Furthermore, the individual level association is evaluated to establish whether the candidate surrogate endpoint can be used to predict the course of the disease in an individual patient. Last but not least, the study level association is investigated to ensure that the treatment effects on the final outcome can be predicted from the effect on the surrogate endpoint.
	Study level association requires data from a number of randomised controlled trials (RCTs) and can be investigated carrying out a bivariate meta-analysis \cite{daniels1997meta,buyse2000validation,bujkiewicz2015uncertainty,lassere2008biomarker}. In this paper we focus on the third level of association only.
	
	A bivariate meta-analytical method that was developed by Daniels \& Hughes \cite{daniels1997meta} can be used to validate a candidate surrogate endpoint, by evaluating the association pattern between the treatment effects on the surrogate and the final outcomes, and to predict treatment effects on the final clinical outcome from the effects on surrogate endpoint.
	This method, implemented in a Bayesian framework, can be used to evaluate a surrogate endpoint in a disease area overall, or in each treatment class separately through a subgroup analysis.  
	
	Traditionally, surrogate relationships between treatment effects on a surrogate endpoint and treatment effects on a final outcome investigated in a disease area using data from all trials regardless of treatment classes or trials of the same class of treatments. For instance, in advance colorectal cancer (aCRC) progression free survival (PFS), tumor response (TR) or time to progression (TTP) have been investigated as potential surrogate endpoints for OS \cite{buyse2007progression,giessen2013progression,ciani2015meta,chirila2012meta}. In previous work, Buyse et al. \cite{buyse2007progression} found a strong association between treatment effects on PFS and OS in this disease area, by including in their meta-analysis studies on one treatment class only (modern chemotherapy). More recently, Ciani et al. \cite{ciani2015meta} investigated the surrogate relationship in aCRC across all modern treatments, including a range of targeted therapies, which led to suboptimal surrogate relationship in this disease area. They concluded that in aCRC the association patterns could vary across treatment classes and a surrogate relationship observed in a specific treatment class may not directly apply across other treatment classes or lines of treatment. This may be particularly important for targeted treatments used only in a subset of population. For example anti-EGFR treatments are recommended for patients without a KRAS/panRAS mutation as these mutations are associated with resistance to the anti-EGFR therapies \cite{KRASmut2017,heterogeneityofCrC2013} and the association pattern might be different for this particular treatment class in this subset of population with this unique characteristic. Furthermore, Giessen et al. \cite{giessen2013progression} who investigated the surrogate relationships in aCRC including all available treatments and subgroups of therapies, inferred that for validation of surrogacy in targeted treatments such as anti-EGFR therapies or anti-VEGF treatments further research is required once more data become available. Consequently, the assumption that a surrogate relationship remains the same across different treatment classes or lines of treatment does not seem reasonable in aCRC, which may be the case in other disease areas. Therefore potential differences in surrogate relationships across classes should be investigated. This can be achieved by performing subgroup analysis using a standard model (e.g. Daniels and Hughes model \cite{daniels1997meta}) or extending the standard model by adding another level to the hierarchical structure of the model for a surrogate relationship accounting for differences between treatment classes. In this paper, we propose two new methods which allow different degrees of borrowing of information for surrogate relationships across treatment classes aiming to obtain estimates of surrogate relationships with higher precision \cite{efron1975,louis1984,louis2000}. 
	The first approach assumes full exchangeability of surrogate relationships exploiting the similarity of surrogate relationships and borrowing information across treatment classes. The second method is able to relax this assumption, by allowing for partial exchangeability \cite{neuenschwander2016robust} of surrogate relationships across treatment classes to avoid excessive borrowing of information from distinctly different treatment classes. In this model, the parameters describing surrogate relationships can be either exchangeable or non-exchangeable giving more flexibility when the assumption of exchangeability is not reasonable.
	
	The modelling techniques were demonstrated using an example in advanced colorectal cancer where the surrogate relationships may vary across treatment classes \cite{ciani2015meta}. To assess models' performance and compare them with subgroup analysis we carried out a simulation study. In the remainder of this paper, we present the standard model in Section \ref{Standard} , the two proposed models are introduced in Section \ref{New}, the results of the simulation study are demonstrated in Section \ref{SimStdy} and the illustrative example as well as the results from its analysis are presented in Section \ref{Application}. The paper concludes with a discussion in Section 7.
	
	\section{Standard surrogacy model}\label{Standard}
	To investigate surrogate relationships within treatment classes using aggregate data, we performed subgroup analysis adopting a standard surrogacy model that was introduced by Daniels and Hughes \cite{daniels1997meta} for the study level evaluation of potential surrogate markers. Equation (\ref{eq:1}) corresponds to the within-study model where $Y_{1i}$, $Y_{2i}$ are the estimates of treatment effects on surrogate endpoint and on the final outcome (for example log odds ratios for TR and log hazard ratio for OS). These effects follow a bivariate normal distribution with $\mu_{1i}$ and $\mu_{2i}$ corresponding to the true treatment effects on the surrogate and the final clinical outcome respectively while, $\sigma_{1i}$, $\sigma_{2i}$ and $\rho_{wi}$ are the within-study standard deviations for both outcomes and the within-study correlations between the treatment effects on the two outcomes for each study $i$.
	
	\begin{gather*}
	\begin{pmatrix}Y_{1i}\\
	Y_{2i}
	\end{pmatrix} \sim  N
	\begin{pmatrix}
	\begin{pmatrix}
	\mu_{1i}\\
	\mu_{2i}
	\end{pmatrix}\!\!&,&
	\begin{pmatrix}
	\sigma_{1i}^{2} & \sigma_{1i}\sigma_{2i}\rho_{wi}  \\
	\sigma_{1i}\sigma_{2i}\rho_{wi} & \sigma_{2i}^{2}
	\end{pmatrix}
	\end{pmatrix}\numberthis \label{eq:1}\\
	\mu_{2i}|\mu_{1i} \sim N(\lambda_{0}+\lambda_{1}\mu_{1i},\psi^{2})\numberthis \label{eq:2} \\
	\end{gather*}
	
	At the between-studies level (\ref{eq:2}), the true effects on the surrogate endpoint $\mu_{1i}$ are modelled as fixed effects, and the true effects on the final outcome $\mu_{2i}$ have linear relationship with the true effects on the surrogate $\mu_{1i}$. This relationship plays a very important role as it can be used to predict $\mu_{2i}$ from known $\mu_{1i}$ in a new study $i$. The parameters $\lambda_{0}$, $\lambda_{1}$, $\psi^{2}$ correspond to the intercept, the slope and the conditional variance of the linear model and measure the shape of the relationship and the strength of association between the treatment effects on the surrogate endpoint and the effects on the final outcome. 
	
	In the Bayesian framework, the Daniels \& Hughes model was implemented by assuming no prior knowledge about surrogate relationship by using vague prior distributions. This allows the data to dominate the posterior distribution even if the dataset is relatively small. The following prior distributions can be used: $\mu_{1i} \sim N(0,a)$, $\lambda_{0} \sim N(0,a)$, $\lambda_{1} \sim N(0,a)$, $\psi \sim N(0,b)I(0,)$, where $N(0,b)I(0,)$ denotes a normal distribution truncated \cite{bland1999measuring} at the mean $\mu=0$ with standard deviation $s=b$. The parameters $a,b$ are chosen to be sufficiently large and depend on the scale of data.
	
	By adapting this method in our research, we applied this standard model to subsets of data that consist of only one class of treatment examining the surrogate relationship of each subgroup separately, taking motivation from similar analyses in clinical trials \cite{berry1990SubgroupAnal,grouin2005Subgroups}. This kind of analysis is very practical when association patterns in a given disease area are different and the treatment classes consist of many studies. By performing subgroup analysis using the standard model, we explored potential differences in the association patterns across treatment classes and use them as a reference for results obtained with the newly developed methods.
	
	\subsection{Criteria for surrogacy}\label{Criteria}
	
	As we mentioned previously, the parameters $\lambda_{0}$, $\lambda_{1}$, $\psi^{2}$ play a very important role, as they are used to evaluate surrogacy. A good surrogate relationship should imply that $\lambda_{1}\ne0$ as slope establishes the association between treatment effects on the surrogate and the final outcome. Subsequently, having $\psi^{2}=0$ implies that $\mu_{2i}$ could be perfectly predicted given $\mu_{1i}$. The parameter $\lambda_{0}$ corresponds to the intercept and is expected to be zero for a good surrogate relationship. This ensures that no treatment effect on the surrogate endpoint will imply no effect on the final outcome. These three criteria proposed by Daniels \& Hughes \cite{daniels1997meta}, will be referred to as surrogacy criteria in the remainder of this paper. A simple way to examine these surrogacy criteria is to check whether or not zero is included in the 95\% credible intervals (CrIs) of $\lambda_{0}$, $\lambda_{1}$ and to compute the Bayes factor for the hypothesis $H_{1}$:\ $\psi^{2}=0$. The model with $\psi^{2}=0$ is a nested model within the standard model \cite{kass1995reference}, so in order to compare these models, Bayes factors can be computed using the Savage Dickey density ratio \cite{verdinelli1995computing}. To implement the Savage Dickey density ratio, proper prior distributions for $\psi$ are needed. In our research a moderately informative half normal prior distribution $N(0,2)I(0,)$ was used for the conditional standard deviation.
	A strong association pattern requires zero to be included in the CrI of $\lambda_{0}$, zero not to be included in the CrI of $\lambda_{1}$ and the Bayes factor of $\psi^{2}$ to be greater than 3.3 \cite{jeffreys1998theory}. In this paper we used the evaluation framework proposed by Daniels and Hughes. However, there are other definitions and criteria for surrogacy in the literature. Detail review of other evaluation frameworks can be found in Lassere et al.\cite{lassere2008biomarker}.
		
	\subsection{Cross-validation}\label{CrossVal}
	One of the main aims of this paper was to explore whether the two hierarchical methods, that we propose in the next section, improve the predictions of treatment effect on the final outcome (by reducing bias and/or uncertainty) compared to subgroup analysis using the standard model. To evaluate this, a cross-validation procedure was carried out. It is a similar to the 'leave-one-study-out' procedure that was described by Daniels \& Hughes \cite{daniels1997meta} and it is repeated as many times as the number of studies in the data set. 
	In a simulated data scenario, this can be used to draw inferences about predicting the true effect on the final endpoint $\mu_{2i}$ in a 'new' study $i$, however, in a real data scenario true effects are unknown and therefore, we can only compare the observed values $Y_{2i}$ with their predicted intervals. 
	For each study $i (i=1,..,N)$, treatment effect on the final endpoint $Y_{2i}$ is omitted and assumed unknown. This effect is then predicted from the observed effect on the surrogate endpoint $Y_{1i}$ and by taking into account the treatment effects on both outcomes from the remaining studies. In a Bayesian framework it can be achieved by performing Markov chain Monte Carlo (MCMC) simulation.
	The mean predicted effect is equal to the true effect $\hat{\mu}_{2i}$ predicted by MCMC simulation and the variance of the predicted effect is equal to $\sigma_{2i}^{2}+var(\hat{\mu}_{2i}|Y_{1i},\sigma_{1i},Y_{1(-i)},Y_{2(-i)})$ where $Y_{1,2(-i)}$ denote the observed treatment effects from the remaining studies without the study that is omitted in $i^{th}$ iteration \cite{daniels1997meta}. We then checked whether the 95\% predictive interval (constructed using the variance) included the observed value of the treatment difference on the final outcome.

	\section{Methods for surrogate endpoint evaluation incorporating aggregate data from different treatment classes}\label{New}
	
	When subgroup analysis is used to investigate the study level surrogate relationships within treatment classes the validation process may fail due to limited data resulting in estimates of the parameters describing surrogate relationships obtained with considerable uncertainty \cite{berry1990SubgroupAnal}. We propose two hierarchical models to investigate surrogate relationships within treatment classes allowing different degrees of borrowing of information about the parameters of interest, as alternative approaches to subgroup analysis with the standard model. These models were developed to investigate the study level association and therefore they can only be applied to aggregate data (e.g logHR or logOR). They allow for the association patterns to vary across classes taking advantage of the attractive statistical properties of exchangeability \cite{efron1975,louis1984,louis2000}. 	
		
	\subsection{Hierarchical model with full exchangeability (F-EX)}\label{F-EX}
	Our first approach extends the standard model accounting for differences in study level surrogacy across different treatment classes \cite{berry2013Hierarchical,thall2003,chugh2009phase}. 
	Similarly as in the standard model, at the within-study level we assume that correlated and normally distributed observed treatment effects $Y_{1ij}$ and $Y_{2ij}$ (e.g logHR or logOR) in each study $i$ estimate the true treatment effects $\mu_{1ij}$ and $\mu_{2ij}$ on the surrogate and final outcomes respectively. In addition, by introducing index $j$ we account for the differences between the classes. 
	
	\begin{gather*}
	\begin{pmatrix}Y_{1ij}\\
	Y_{2ij}
	\end{pmatrix} \sim  N
	\begin{pmatrix}
	\begin{pmatrix}
	\mu_{1ij}\\
	\mu_{2ij}
	\end{pmatrix}\!\!,&
	\begin{pmatrix}
	\sigma_{1ij}^{2} & \sigma_{1ij}\sigma_{2ij}\rho_{wij} \\
	\sigma_{1ij}\sigma_{2ij}\rho_{wij} & \sigma_{2ij}^{2}
	\end{pmatrix}
	\end{pmatrix}\\
	\mu_{2ij}|\mu_{1ij} \sim N(\lambda_{0j}+\lambda_{1j}\mu_{1ij},\psi_{j}^{2})\numberthis \label{eq:3}\\
	\lambda_{0j} \sim N(\beta_{0},\xi_{0}^{2}), \lambda_{1j} \sim N(\beta_{1}, \xi_{1}^{2})\\
	\newline 
	\end{gather*}
	The parameters $\sigma_{1ij}$, $\sigma_{2ij}$, $\rho_{wij}$ correspond to the within-study variances and within-study correlations for each study $i$ in treatment class $j$. The observed estimates $Y_{1ij}$, $Y_{2ij}$, $\sigma_{1ij}$, $\sigma_{2ij}$ are aggregate data extracted from systematic review RCTs whilst, the within-study correlations $\rho_{wij}$ can be calculated using a bootstrapping method from individual patient data (IPD).
	Similarly as in the standard model, the true effects $\mu_{1ij}$  on the surrogate endpoint are modelled as fixed effects. 
	
	In contrast to the standard surrogacy model, this method assumes unique surrogate relationships between true treatment effects on the surrogate endpoint and the final outcome across treatment classes in a single model, allowing for borrowing of information across them.
	Each relationship between the true effects on the surrogate endpoint $\mu_{1ij}$ and the final outcome $\mu_{2ij}$ is described by a linear model where, $\lambda_{0j}$ denotes the intercept of the $j^{th}$ treatment class and $\lambda_{1j}$ establishes the relationship between treatment effects on surrogate and final outcomes within the treatment class $j$. To evaluate whether a candidate endpoint is considered a valid surrogate endpoint in a given treatment class, all three surrogacy criteria need to be met for this particular class.
	Implementing this model in the Bayesian framework, we place non-informative prior distributions on the model parameters such as: $\beta_{0}, \beta_{1} \sim N(0,a)$ and $\xi_{0},\xi_{1} \sim N(0,b)I(0,)$, $\mu_{1ij} \sim N(0,a)$ and $\psi_{j}\sim N(0,b)I(0,)$. Similarly as in the standard model $a,b$ are chosen to be sufficiently large and depend on the scale of data.
	
	F-EX model extends the standard model (described in section \ref{Standard}) by including an additional layer of hierarchy to the linear relationship between true effects on the surrogate and the final outcome, assuming that slopes and intercepts are fully exchangeable across treatment classes. This can be implemented by placing common normal distributions on $\lambda_{0j}$ and $\lambda_{1j}$ with means and variances $\beta_{0}$, $\xi_{0}^{2}$ and $\beta_{1}$, $\xi_{1}^{2}$ leading to borrowing of information across treatment classes. Hierarchical models have desirable statistical properties that allow us to improve our inferences taking advantage of borrowing of information from other treatment classes. The exchangeable estimates, however, are shrunk towards the means $\beta_{0}$, $\beta_{1}$ and the amount of shrinkage depends on the number of studies within each class, the between treatment class heterogeneity \cite{neuenschwander2016robust} and the number of treatment classes. Although these statistical properties are very attractive in terms of potential reduction of uncertainty around the parameters of interest, they are advantageous only when the assumption of exchangeability is reasonable, otherwise there is a danger of excessive shrinkage.

	\subsection{Hierarchical model with partial exchangeability (P-EX)}\label{P-EX}
	F-EX method can be extended into a method with partial exchangeability similarly as the method proposed by Neuenschwander et al. \cite{neuenschwander2016robust}. This model is able to relax the assumption of exchangeability allowing the parameters of interest for each class to be either exchangeable with all or some of the parameters from other treatment classes or non-exchangeable. The proposed method is more flexible compared to F-EX model, in particular in data scenarios where the assumption of exchangeability is not reasonable for some of the treatment classes.
	
	The within and the between study level of this model is exactly the same as in the method with full exchangeability where, $Y_{1ij}$, $Y_{2ij}$ are the treatment effects on the surrogate and final clinical outcomes and they follow a bivariate normal distribution with mean values corresponding to the true treatment effects $\mu_{1ij}$ and $\mu_{2ij}$ on the two outcomes.  
	
	\begin{gather*}
	\begin{pmatrix}Y_{1ij}\\
	Y_{2ij}
	\end{pmatrix} \sim  N
	\begin{pmatrix}
	\begin{pmatrix}
	\mu_{1ij}\\
	\mu_{2ij}
	\end{pmatrix}\!\!,&
	\begin{pmatrix}
	\sigma_{1ij}^{2} & \sigma_{1ij}\sigma_{2ij}\rho_{wij} \\
	\sigma_{1ij}\sigma_{2ij}\rho_{wij} & \sigma_{2ij}^{2}
	\end{pmatrix}
	\end{pmatrix}\\
	\mu_{2ij}|\mu_{1ij} \sim N(\lambda_{0j}+\lambda_{1j}\mu_{1ij},\psi^{2}_{j})\\
	\lambda_{0j}\sim N(\beta_{0},\xi_{0}^{2})\numberthis \label{eq:4}\\
	\lambda_{1j}=\left\{\fontsize{10}{25}
	\begin{array}{cc}
	\lambda_{1j} \sim N(\beta_{1},\xi_{1}^{2}) & \quad\text{if } p_{j} = 1 \\
	\lambda_{1j} \sim N(0,b) & \quad\text{if } p_{j} = 0 
	\end{array}
	\right.
	\newline
	\end{gather*}
	
	However, the parameters of slopes are modelled in a different way compared to F-EX model. In this approach two possibilities arise for these parameters for each treatment class $j$. When $p_{j}=1$ the parameter $\lambda_{1j}$ can be exchangeable with some or all the parameters of the slopes from the other treatment classes via an exchangeable component. It follows a common normal distribution with other slopes as in F-EX model.
	On the other hand, when $p_{j}=0$ the slope can be non-exchangeable with any slopes from the other treatment classes. In this case a vague prior distribution can be placed on the parameter, as in the standard model. The method evaluates the degree of borrowing of information for each parameter $\lambda_{1j}$ by using these two components with respective mixture weights.
	
	The main advantage of this method is that it allows the mixture weights to be inferred from the data. In each MCMC iteration, the sampler chooses between the two components by using a Bernoulli distribution $p_{j}\sim Bernoulli(\pi_{j})$. By calculating the posterior mean of this Bernoulli distribution we derive the mixture weights of each treatment class.
	The hyper-parameters $\pi_{j}$ of the Bernoulli prior distribution can be either fixed or, in a fully Bayesian framework, they can follow  a prior distribution for example, a Beta distribution $\pi_{j}\sim Beta(1,1)$. We have used fixed $\pi_{j}$, since placing a prior distribution required longer chains to converge and provided almost the same results.
	
	In a special case where $p_{j}=1$ for all treatment classes, P-EX model reduces to full exchangeability model as it uses only the exchangeable component. Having $p_{j}=0$ for all treatment classes makes the P-EX model equivalent to subgroup analysis using the standard model as only the non-exchangeable component is used to estimate $\lambda_{1j}$ in this case. 
	In a Bayesian framework vague prior distributions can be placed on the parameters $\beta_{0},\beta_{1}$, $\xi_{0},\xi_{1}$, $\mu_{1ij}$ as in F-EX model.

	\section{Software Implementation and computing}
	All models were implemented in OpenBUGS \cite{R2w} where posterior estimates were obtained using MCMC simulations performing 50000 iterations (after discarding 20000 iterations as burn-in period). The OpenBUGS code of F-EX and P-EX models can be found in the supplementary material (sections D.3, D.4). Convergence was assessed visually by checking the history, chains and autocorrelation plots using graphical tools in OpenBUGS and R. All estimates are presented as means with corresponding 95\% CrIs. The median was used only for the estimates of the conditional variances as a measure of central tendency since their posterior distributions were very skewed. The cross-validation procedure was performed in R using R2OpenBUGS \cite{R2w} package to execute OpenBUGS code multiple times.

	\section{Simulation study}\label{SimStdy}
	The proposed hierarchical methods allow different levels of borrowing of information for the parameters of interest. F-EX model assumes exchangeability for slopes whilst, the P-EX model allows for partial exchangeability for these parameters. We carried out a simulation study to assess the performance of the hierarchical methods and to compare them with subgroup analysis conducted using the standard model. We evaluated the performance of the methods in distinct data scenarios generated assuming different strengths of association within classes, different levels of similarity of the association patterns across classes and different number of studies per class. We evaluated the models' ability to identify treatment classes with strong association patterns and to make predictions of the treatment effect on the final outcome in a new study from a treatment effect measured on the surrogate endpoint.

	\subsection{Data generation process and scenarios}\label{GenerationPr}
	We simulated data under nine different scenarios generating 1000 replications for each scenario. Each replication included average treatment effects on the surrogate and the final outcome (and corresponding standard errors (SEs) and within-study correlations) from a number of studies of treatments belonging to five treatment clases.	
	We assumed that the data in each treatment class had a different heterogeneity pattern. Therefore, to have a control over such heterogeneity patterns when simulating the data we needed to make an assumption about the distribution of the true effects both on the surrogate and the final endpoints. The standard model by Daniels \& Hughes assumes fixed effect for the true effects on the surrogate endpoint (no common distribution) making difficult to control the heterogeneity patterns when simulating the data. To avoid this issue, we simulated data using a product normal formulation of bivariate random effect meta-analysis (BRMA) (equation set (\ref{eq:5})), assuming normal random effects on the surrogate endpoint.
	Apart from this assumption, this method is the same as Daniels \& Hughes model using a bivariate normal distribution to describe the within-study variability and a linear relationship to model the association between the surrogate and the final outcome. 
		
	\begin{gather*}
	\begin{pmatrix}Y_{1ij}\\
	Y_{2ij}
	\end{pmatrix} \sim  N
	\begin{pmatrix}
	\begin{pmatrix}
	\mu_{1ij}\\
	\mu_{2ij}
	\end{pmatrix}\!\!,&
	\begin{pmatrix}
	\sigma_{1ij}^{2} & \sigma_{1ij}\sigma_{2ij}\rho_{wij} \\
	\sigma_{1ij}\sigma_{2ij}\rho_{wij} & \sigma_{2ij}^{2}
	\end{pmatrix}
	\end{pmatrix}\\
	\mu_{1ij}\sim N(\eta_{1j},\psi_{1j}^{2})\numberthis \label{eq:5}\\
	\mu_{2ij}|\mu_{1ij} \sim N(\eta_{2ij},\psi_{2j}^{2})\\
	\eta_{2ij} = \lambda_{0j}+\lambda_{1j}\mu_{1ij}\\
	\psi_{1j} = \frac{\psi_{2j}}{|\lambda_{1j}|\sqrt{(1/\rho_{bj}^2)-1}}
	\end{gather*}
	Simulating data from this model, however, can lead to results obtained with increased uncertainty, as the models used to analyse the data make fewer distributional assumptions.
		
	To generate the data, we pursued the following steps:
	\begin{enumerate}
		\item Set the number of classes $N=5$
		
		\item Simulate the data for each class separately using BRMA model (eq. \ref{eq:5}) under 3 main designs.				
		
		\item Create three sets of scenarios: two with fixed number of studies ($n_{j}$=16 and $n_{j}$=8, j=1,...,5) per treatment class and one with unbalanced classes ($n_{1}$ = 4, $n_{2}$ = 8, $n_{3}$ = 6, $n_{4}$ = 10, $n_{5}$ = 7 ). We applied the three sets of scenarios to each design. In total, we have  9 scenarios (3 designs $\times$ 3 sets = 9 scenarios).

		\item Simulate the true effects using model (eq. (5))
	\end{enumerate}

	The values of the parameters are listed in Table 1 and a short description of each design can be found below: 
	\newline \newline
		\textbf{Design 1:}\newline 	
	In the first design, our aim was to illustrate the properties of exchangeability. We simulated data in five treatment classes assuming high degree of similarity for their slopes and intercepts. The data in each treatment class were simulated assuming strong association (see surrogacy criteria in Section \ref{Criteria} ) for each individual class but weak overall. 
	\newline\newline
	\textbf{Design 2:}\newline
	The second design illustrates the case where there is a treatment class with very different association pattern (slope) compared to the other classes. This implies that the assumption of exchangeability is in doubt for this parameter in this particular class. Similarly as in the first scenario, we assumed strong association for each individual class. 
	\newline\newline
	\textbf{Design 3:}\newline	
	The last design focuses on the association patterns of strengths that vary across treatment classes, investigating whether the proposed methods can estimate a strong association pattern better compared to subgroup analysis with the standard model and whether they can distinguish between the different association patterns despite borrowing of information across treatment classes. To achieve this, we generated three out of five treatment classes with strong association and the remaining two classes with a weak association.
	
		\begin{table}[h!]
		\label{Design}
		\caption{Simulation designs}
		\centering
		\begin{tabular}{ccc}
			\multicolumn{1}{c}{1$^{st}$ design} &\multicolumn{1}{c} {2$^{nd}$ design}&\multicolumn{1}{c} {3$^{rd}$ design}\\
			\hline
			$\lambda_{11} = 0.40$, $\rho_{b1} = 0.89$&$\lambda_{11} = 0.60$, $\rho_{b1} = 0.93$&$\lambda_{11} = 0.40$, $\rho_{b1} = 0.90$\\  
			$\lambda_{12} = 0.45$, $\rho_{b2} = 0.90$&$\lambda_{12} = 1.55$, $\rho_{b2} = 0.99$&$\lambda_{12} = 0.50$, $\rho_{b2} = 0.70$\\ 
			$\lambda_{13} = 0.50$, $\rho_{b3} = 0.91$&$\lambda_{13} = 1.60$, $\rho_{b3} = 0.99$&$\lambda_{13} = 0.60$, $\rho_{b3} = 0.93$\\ 
			$\lambda_{14} = 0.55$, $\rho_{b4} = 0.92$&$\lambda_{14} = 1.65$, $\rho_{b4} = 0.99$&$\lambda_{14} = 0.70$, $\rho_{b4} = 0.75$\\
			$\lambda_{15} = 0.60$, $\rho_{b5} = 0.93$&$\lambda_{15} = 1.70$, $\rho_{b5} = 0.99$&$\lambda_{15} = 0.80$, $\rho_{b5} = 0.95$\\
			$\lambda_{0j} = 0$&$\lambda_{0j} = 0$&$\lambda_{0j} = 0$\\
			$\sigma_{1ij,2ij}=0.1$&$\sigma_{1ij,2ij}=0.1$&$\sigma_{1ij,2ij}=0.1$\\
			$\rho_{wij} = 0.4$&$\rho_{wij} = 0.4$&$\rho_{wij} = 0.4$\\
			$\psi_{2j} = 0.08$&$\psi_{2j} = 0.08$&$\psi_{21,23,25} = 0.08$\\
			&& $\psi_{22,24} = 0.30$\\
			$\eta_{1j} = 0.3$&$\eta_{1j} = 0.3$&$\eta_{1j} = 0.3$\\
			\hline
		\end{tabular}
	\end{table}	
\newpage
	\subsection{Performance measures}\label{Perfmeasures}
	 To evaluate the goodness of fit of the models, we calculated the coverage probability of the 95\% CrIs of $\lambda_{1j}$ and the 95\% predictive intervals of $\mu_{2ij}$. The absolute bias and the root mean square error (RMSE) of $\hat{\lambda}_{1j}$ and $\hat{\mu}_{2ij}$ were also monitored and reported in the tables. In order to investigate potential decrease in the degree of uncertainty of the estimates as a result of borrowing of information across treatment classes, we calculated ratios of the width of the 95\% CrIs. The width ratio $w_{\lambda_{1j}^{FEX,(PEX)}}$/$w_{\lambda_{1j}^{subgr}}$ was defined as the ratio of the widths of the CrIs of $\lambda_{1j}$ from F-EX or P-EX to the width of the CrIs of $\lambda_{1j}$ from subgroup analysis using the standard model. Similarly, the width ratio $w_{\mu_{2ij}^{FEX,(PEX)}}$/$w_{\mu_{2ij}^{subgr}}$ was the ratio of the 95\% predictive intervals of the true effects $\mu_{2ij}$ from F-EX or P-EX to the width of the predictive intervals of $\mu_{2ij}$ from subgroup analysis using the standard model. We also monitored the largest Monte Carlo error (MCE) of the simulations as an index of accuracy of the Monte Carlo samples.
	 
	 Furthermore, a cross-validation procedure was applied to each method across the simulated data scenarios. In the simulation study, the true effect on the final endpoint $\mu_{2ij}$ was known, since it had been simulated ,therefore the cross-validation procedure was applied on the true effects (in real data scenarios we compare the predicted effect with the observed effect) by checking whether the simulated value of the true effect $\hat\mu_{2ij}$ was included in the predictive interval of $\mu_{2ij}$.	 

	\subsection{Results}
	All the tables in the results section list the performance of the posterior means of $\hat{\lambda}_{1j}$, the performance of the posterior means of $\hat{\mu}_{2ij}$ as well as the probabilities of estimating a strong association pattern  (see definition in Section \ref{Criteria}) for each class across methods. 
	The following section presents the results of the analysis by reporting the coverage probabilities of the CrIs of $\lambda_{1j}$ and $\mu_{2ij}$ for each scenario (by taking the mean of coverage probabilities across classes), the overall absolute bias and RMSE of $\hat{\lambda}_{1j}$ and $\hat{\mu}_{2ij}$, the width ratios of $\lambda_{1j}$ and $\mu_{2ij}$ for each scenario (by calculating the mean of the width ratios of $\lambda_{1j}$ across classes and the mean of the width ratios of $\mu_{2ij}$ across studies and classes), the MCE and the probability to estimate a strong association pattern by fitting each model.
	Detailed results for the performance of $\hat{\lambda}_{1j}$ and $\hat{\mu}_{2ij}$ for each class separately and across methods are listed in the Supplementary material (see section B and section C).
	
	\subsubsection{Performance of the estimates $\hat{\lambda}_{1j}$}
	Table \ref{Performancelambda1} presents the results across the nine scenarios reporting averages of the measures we monitored for $\hat{\lambda}_{1j}$ over the five classes of treatment. The performance of the models varied in terms of the coverage probability of the 95\% CrIs of $\lambda_{1j}$ across scenarios. In the scenarios 1, 4 and 7 where the number of studies per class was relatively high, the models achieved 95\% coverage probabilities. However, in the scenarios where the number of studies was smaller the coverage probability was higher due to increased uncertainty and likely to the fact that the model we used in the generation process was slightly different from models used to fit the data.								
    Monte Carlo errors were small across most of the scenarios implying good accuracy of the Monte Carlo samples and that convergence was achieved in those scenarios across all the methods. However, in scenarios where the data were limited (scenarios 3, 6, 9) subgroup analysis with the standard model yielded larger MCEs. This implies that subgroup analysis requires longer chains to achieve the same level of convergence as the other two models.
    
    In the first three scenarios (1st design), where the treatment classes were very similar in terms of patterns (similar slopes), F-EX and P-EX were superior compared to subgroup analysis as they gave posterior means of slopes with lower absolute bias,  RMSE and reduced uncertainty (narrower 95\% CrIs) due to borrowing of information across classes. P-EX model achieved almost the same level of borrowing of information as F-Ex model, with mixtures weights were very close to 1 across treatment classes (see details in the supplementary material D.1 where the mixture weights are listed). Overall, the proposed hierarchical models performed better compared to subgroup analysis but the difference was more pronounced in the scenarios with small number of studies.
	In the second design (scenarios 4, 5, 6), where the exchangeability assumption was not reasonable for one of the classes, P-EX model yielded the most robust results. The model resulted in the posterior means with the smallest absolue bias and RMSE, reducing the degree of borrowing of information for the class with the distinctly different (the mixture weights in this class were $p_{1}=0.56$, $p_{1}=0.31$ and $p_{1}=0.80$ respectively) whilst still borrowing information across the remaining classes ($p_{2},p_{3},p_{4},p_{5}\approx 0.97$). On the other hand, F-EX performed poorer compared to the other methods in scenario 6 with unbalanced and relatively small number of studies per class, leading to more biased results. This indicates that F-EX model is not appropriate when the assumption of exchangeability is not reasonable. Subgroup analysis using the standard model achieved decent performance only in the forth scenario where there were sufficient data. In the third design (scenarios 7, 8, 9) the proposed models achieved superior performance compared to subgroup analysis for the estimates of $\lambda_{1j}$, similarly as in the first three scenarios.
	
	The last column of Table \ref{Performancelambda1} shows the probabilities of estimating a strong association pattern across the data scenarios and models. F-EX and P-EX methods estimated the surrogacy (based on the three surrogacy criteria) better compared to subgroup analysis across all scenarios.
	In the first design (scenarios 1, 2, 3) where the association was designed to be strong for all the classes, F-EX and P-EX models predicted a strong association pattern in more than 85\% of the simulations. Subgroup analysis predicted the 81\% of them in the 1st scenario but its performance reduced noticeably in the 2nd and 3rd scenario where the data were more sparse. In the second design (scenarios 4, 5, 6) with strong association patterns across all classes, P-EX and F-EX estimated more than 87\% of the association patterns across these three scenarios. Subgroup analysis performed well only in the 4th scenario predicting the 89\% of the association patterns but its performance gradually reduced as the number of studies was decreased in scenario 5 and 6. 
	
	\begin{table}[h!]
	\centering
	\fontsize{9}{14}\selectfont
	\caption{Performance of $\hat{\lambda}_{1j}$}
	\label{Performancelambda1}
	\begin{tabular}{lllcccccc}   
		\hline
		Scenario&\shortstack{No of studies\\across classes}& Methods   & \shortstack{Coverage\\(Mean)}  &\shortstack{Absolute\\Bias\\(Mean)}&RMSE& \shortstack{\\Width Ratio \\ (Mean)}&MCE&\shortstack{\\Prob of\\ strong association\\(Mean)}\\
		\hline
		&$\textbf{1st Design}$&&&&&&&\\
		1st&Fixed ($n_j=16$) &Subgroup Analysis& 0.95&0.08 & 0.10  &     &0.003    &0.81   \\
		&                    &F-EX model       & 0.95&0.06 & 0.07  & 0.72&0.002&0.85  \\
		&                    &P-EX model       & 0.96&0.06 & 0.07  & 0.72&0.002&0.85\\
		&&&&&&&&\\
		2nd&Fixed ($n_j=8$)&Subgroup Analysis  & 0.98&0.11 & 0.15  & &0.005&0.71     \\
		&                  &F-EX model         & 0.97&0.07 & 0.09  & 0.60&0.003&0.89  \\
		&                  &P-EX model         & 0.97&0.07 & 0.09  & 0.61&0.003&0.90\\
		&&&&&&&&\\
		3rd&Unbalanced                  &Subgroup Analysis    & 0.99&0.13 &0.18   & &0.017&0.56      \\
		&($n_{1}=4$,$n_{2}=8$,$n_{3}=6$,&F-EX model           & 0.99&0.07 &0.09   & 0.52&0.003&0.89  \\
		&$n_{4}=10$,$n_{5}=7$)          &P-EX model           & 0.99&0.07 &0.09   & 0.53&0.004&0.88\\
		\hline
		&$\textbf{2nd Design}$&&&&&&&\\
		4th&Fixed ($n_j=16$)&Subgroup Analysis    & 0.95&0.09 & 0.11 &      &0.007&0.89\\
		&                   &F-EX model           & 0.94&0.08 & 0.10 & 0.90&0.005&0.91   \\
		&                   &P-EX model           & 0.94&0.07 & 0.09 & 0.86&0.004&0.91\\
		&&&&&&&&\\
		5th&Fixed ($n_j=8$)&Subgroup Analysis    & 0.97&0.14 & 0.17 &      &0.007&0.88   \\
		&                  &F-EX model           & 0.96&0.12 & 0.15 & 0.86&0.005&0.92  \\
		&                  &P-EX model           & 0.97&0.10 & 0.12 & 0.78&0.005&0.92\\ 
		&&&&&&&&\\
		6th&Unbalanced                  &Subgroup Analysis    & 0.98&0.15 &0.20 & &0.025&0.72   \\
		&($n_{1}=4$,$n_{2}=8$,$n_{3}=6$,&F-EX model           & 0.96&0.17 &0.21 & 0.70&0.011&0.88  \\
		&$n_{4}=10$,$n_{5}=7$)          &P-EX model           & 0.97&0.14 &0.18 & 0.70&0.011&0.87\\
		\hline
		&$\textbf{3rd Design}$&&&&&&&\\
		7th&Fixed ($n_j=16$)&Subgroup Analysis    & 0.95&0.11 &0.14   & &0.003&   \\
		&                   &F-EX model           & 0.95&0.09 &0.11   & 0.79&0.002&  \\
		&                   &P-EX model           & 0.95&0.09 &0.11   & 0.79&0.003&\\
		&&&&&&&&\\
		8th&Fixed ($n_j=8$)&Subgroup Analysis    & 0.97&0.17 &0.22   & &0.006&   \\
		&                  &F-EX model           & 0.96&0.11 &0.14   & 0.67&0.004&  \\
		&                  &P-EX model           & 0.96&0.11 &0.14   & 0.67&0.004&\\
		&&&&&&&&\\
		9th&Unbalanced                   &Subgroup Analysis    & 0.98&0.19 &0.25 & &0.021&   \\
		&($n_{1}=4$,$n_{2}=8$,$n_{3}=6$, &F-EX model           & 0.97&0.12 &0.15 & 0.56&0.005&  \\
		&$n_{4}=10$,$n_{5}=7$)           &P-EX model           & 0.97&0.12 &0.15 & 0.57&0.005&\\
		\hline
	\end{tabular}  
\end{table}

	\newpage

	Table \ref{surrogacy} presents the results from the last 3 scenarios (3rd design), where the surrogate relationships varied across classes. F-EX and P-EX methods were able to estimate a strong association pattern with higher probability compared to subgroup analysis in the classes where the association was designed to be strong. At the same time, the methods successfully identified classes with strong association patterns from a mixture of classes with weak and strong association patterns, even for the scenarios with relatively few studies per class where subgroup analysis failed almost completely to identify.
	The probabilities of estimating a strong association per class in designs 1 and 2 are presented in the Supplementary material (see section B.1, B.2, B.3, B.4, B.5, B.6). 
	
	\begin{table}[h!]
			\caption{Probabilities of estimating a strong association pattern per class in the 3rd design}
			\label{surrogacy}
			\fontsize{9}{15}\selectfont\label{key}
			\centering
			\begin{tabular}{lllccc}
				\hline
				Scenario&\shortstack{No of studies\\across classes}&$\textbf{Treatment classes}$  &Subgroup Analysis&F-EX model&P-EX model  \\
				\hline 
7th&Fixed ($n_{j}=16$)&1$^{st}$ class & 0.82 &  0.84 &0.84  \\
				&&2$^{nd}$ class$^*$ & 0.00 &  0.00 &0.00   \\
				&&3$^{rd}$ class & 0.83 &  0.85 &0.85   \\
				&&4$^{th}$ class$^*$ & 0.00 &  0.00 &0.00   \\
				&&5$^{th}$ class & 0.80 &  0.80 &0.80  \\
				\hline 
8th&Fixed ($n_{j}=8$)&1$^{st}$ class & 0.78 &  0.89 &0.89  \\
				 &&2$^{nd}$ class$^*$ & 0.04 &  0.05 &0.05   \\
				 &&3$^{rd}$ class & 0.80 &  0.90 &0.90   \\
				 &&4$^{th}$ class$^*$ & 0.06 &  0.06 &0.06   \\
				 &&5$^{th}$ class & 0.85 &  0.87 &0.86  \\    		
				\hline
9th&Unbalanced	&1$^{st}$ class & 0.06 &  0.82 &0.80  \\
&($n_{1}=4$,$n_{2}=8$,&2$^{nd}$ class$^*$ & 0.06 &  0.07 &0.07   \\
&$n_{3}=6$,$n_{4}=10$,&3$^{rd}$ class & 0.65 &  0.91 &0.91   \\
&$n_{5}=7$)		&4$^{th}$ class$^*$ & 0.03 &  0.03 &0.03   \\
&				&5$^{th}$ class & 0.82 &  0.89 &0.89  \\    		
				\hline 
			\multicolumn{6}{l}{*Treatment classes with weak association pattern}	   			
			\end{tabular} 		
	\end{table}

	\subsubsection{Performance of predictions $\hat{\mu}_{2ij}$}
	
	Table \ref{PerformancePred} shows the results from cross-validation procedure which resulted in the posterior means ($\hat{\mu}_{2ij}$) and 95\% predictive intervals of the true effects $\mu_{2ij}$. It presents the same measures as Table \ref{Performancelambda1} averaged over the five classes. In scenarios 1, 4 and 7, the models achieved 95\% coverage due to the large amount of data, however, in the remaining scenarios where the number of studies was smaller the models yielded higher coverages probabilities. F-EX and P-EX had small Monte Carlo errors across all scenarios, however, subgroup analysis gave on average significantly larger MCEs compared to the proposed methods in scenarios 3, 6, 9. More specifically, in the treatment classes where the surrogate relationship was designed to be weak or the data were limited the MCEs were higher (see details in the supplementary material C.3, C.6, C.9). This indicates that subgroup analysis with the standard model requires longer chains for its posteriors to achieve the same level of convergence as the other two methods.
	
	In the first three scenarios, F-EX and P-EX models outperformed subgroup analysis in terms of the absolute bias, RMSE and the uncertainty of $\hat{\mu}_{2ij}$. However, there was no winner between them as both methods had almost the same degree of borrowing of information resulting in 7\%, 20\%, and 33\% narrower predictive intervals compared to subgroup analysis across these three scenarios respectively. 
	In the scenarios 4 ,5 ,6 (2nd desing), P-EX yielded posterior means with the smallest absolute bias, RMSE and CrIs with the smallest width ratio across classes. Furthermore, P-EX method gave the most robust results for the 'extreme' treatment class relaxing 44\%, 70\% and 20\% the borrowing of information in this class across the scenarios (see the mixture weights in the supplementary material D.1). In the 6th scenario F-EX performed poorer compared to P-EX model leading to biased results especially for the treatment class where the surrogacy was different and the exchangeability assumption unreasonable (1st class in the section B.6 of the supplementary material). Subgroup analysis performed almost equally well as the P-EX model in the 4th scenario where the number of studies per class was relatively large.		
	The last three scenarios (3rd design) gave similar results as the first three in terms of the uncertainty, the absolute bias and the RMSE of $\hat{\mu}_{2ij}$. F-EX P-EX models performed equally well, whilst subgroup analysis with the standard model was the worst approach resulting in inflated predictive intervals, larger RMSE and worse MCE in all cases.

	\begin{table}[h!]
	\centering
	\fontsize{9}{14}\selectfont
	\caption{Performance of $\hat{\mu}_{2ij}$}
	\label{PerformancePred}
	\begin{tabular}{lllccccc}   
		\hline
		Scenario&\shortstack{No of studies\\across classes}& Methods   & \shortstack{Coverage\\(Mean)}  &\shortstack{Absolute\\Bias\\(Mean)}&RMSE& \shortstack{\\Width Ratio \\ (Mean)}&MCE\\
		\hline
		&$\textbf{1st Design}$&&&&&&\\
		1st&Fixed ($n_j=16$) &Subgroup Analysis& 0.95&0.09 & 0.11  &     &0.003   \\
		&                    &F-EX model       & 0.95&0.08 & 0.10  & 0.93&0.002  \\
		&                    &P-EX model       & 0.95&0.08 & 0.10  & 0.93&0.002\\
		&&&&&&&\\
		2nd&Fixed ($n_j=8$)&Subgroup Analysis  & 0.98&0.11 & 0.13 &     &0.010   \\
		&                  &F-EX model         & 0.98&0.08 & 0.10 & 0.80&0.004  \\
		&                  &P-EX model         & 0.98&0.08 & 0.10 & 0.80&0.004\\
		&&&&&&&\\
		3rd&Unbalanced                  &Subgroup Analysis    & 0.99&0.12 &0.18   &     &0.023   \\
		&($n_{1}=4$,$n_{2}=8$,$n_{3}=6$,&F-EX model           & 0.99&0.08 &0.11   & 0.67&0.005  \\
		&$n_{4}=10$,$n_{5}=7$)          &P-EX model           & 0.99&0.09 &0.11   & 0.68&0.008\\
		\hline
		&$\textbf{2nd Design}$&&&&&&\\
		4th&Fixed ($n_j=16$)&Subgroup Analysis    & 0.95&0.13 & 0.18  & &0.009       \\
		&                   &F-EX model           & 0.95&0.13 & 0.18  & 0.97&0.008  \\
		&                   &P-EX model           & 0.96&0.12 & 0.17  & 0.96&0.008\\
		&&&&&&&\\
		5th&Fixed ($n_j=8$)&Subgroup Analysis    & 0.99&0.16 & 0.20 &     &0.015   \\
		&                  &F-EX model           & 0.98&0.15 & 0.19 & 0.92&0.009  \\
		&                  &P-EX model           & 0.98&0.14 & 0.18 & 0.87&0.008\\ 
		&&&&&&&\\
		6th&Unbalanced                  &Subgroup Analysis    & 0.99&0.18 &0.23   & &0.021       \\
		&($n_{1}=4$,$n_{2}=8$,$n_{3}=6$,&F-EX model           & 0.99&0.18 &0.22   & 0.80&0.009  \\
		&$n_{4}=10$,$n_{5}=7$)          &P-EX model           & 0.99&0.15 &0.19   & 0.77&0.010\\
		\hline
		&$\textbf{3rd Design}$&&&&&&\\
		7th&Fixed ($n_j=16$)&Subgroup Analysis    & 0.95&0.16 &0.23   & &0.006       \\
		&                   &F-EX model           & 0.95&0.16 &0.22   & 0.96&0.004  \\
		&                   &P-EX model           & 0.95&0.16 &0.22   & 0.96&0.004\\
		&&&&&&&\\
		8th&Fixed ($n_j=8$)&Subgroup Analysis    & 0.98&0.18 &0.26 & &0.017       \\
		&                  &F-EX model           & 0.97&0.16 &0.22 & 0.85&0.006  \\
		&                  &P-EX model           & 0.97&0.16 &0.22 & 0.85&0.006\\
		&&&&&&&\\
		9th&Unbalanced                   &Subgroup Analysis    & 0.98&0.20 &0.28 & &0.027       \\
		&($n_{1}=4$,$n_{2}=8$,$n_{3}=6$, &F-EX model           & 0.97&0.17 &0.21 & 0.72&0.008  \\
		&$n_{4}=10$,$n_{5}=7$)           &P-EX model           & 0.97&0.17 &0.21 & 0.72&0.009\\
		\hline
	\end{tabular}  
\end{table}

	\subsection{Discussion of the results}
	
	The aim of the simulation study was to illustrate and assess the performance of the methods under different scenarios. The models gave 95\% coverage probabilities in the scenarios 1, 4, and 7 where the number of studies was sufficiently large (16 for each class). However, in the remaining scenarios the coverage probabilities were higher than 95\%, which means that the methods derived more conservative CrIs of parameters than expected. This is largely due to the sparsity of the data in these scenarios but may also be partly due to different models being used to simulate and analyse the data as explained in section \ref{GenerationPr}.
	In the first design (scenarios 1, 2, 3) where the assumption of exchangeability was reasonable, F-EX and P-EX models performed better than the subgroup analysis giving on average narrower 95\% CrIs of $\lambda_{1j}$ and 95\% predictive intervals of $\mu_{2ij}$. This indicates that P-EX model successfully identified the correct level of borrowing of information inferring that the mixture weights should be very close to 1. P-EX model was the best choice in all the scenarios of the second design (scenarios 4, 5, 6) where there was a treatment class with distinctly different slope. It relaxed the degree of borrowing of information for the 'extreme' treatment class, giving the most accurate posterior means of the slopes.
	Moreover, P-EX model was the best choice in terms of predictions of the true effect on the final endpoint, reducing the width of predictive intervals by 4\%, 13\% and 23\% compared to subgroup analysis in each scenario respectively. Last but not least, the proposed methods estimated the strong association patterns better compared to subgroup analysis across all data scenarios. In particular, in scenarios 3, 6 and 9, where the data were sparse, the proposed hierarchical methods were able to estimate surrogacy significantly better compared to the subgroup analysis. This illustrates well the benefits of using hierarchical methods when data are limited. Furthermore, as illustrated by scenarios 7,8 and 9, F-EX and P-EX could easily distinguish between the different association patterns as they identified treatment classes with strong association patterns and at the same time did not overestimate the strength of the association in the classes where the association was designed to be weak.

	\section{Application: Advanced colorectal cancer}\label{Application}
	\subsection{Data}
	We illustrate the proposed methodology in an example in aCRC. The data were obtained from a systematic review conducted by Ciani et al. \cite{ciani2015meta} which included 101 RCTs published between 2003 and 2013, evaluating multiple interventions in aCRC. The review consist of trials that report treatment effects on OS or/and on alternative endpoints such as progression-free-survival or tumor response (PFS, TR). OS was defined as the time from randomisation to time of death, PFS was set as the time from randomisation to tumor progression or death from any cause. Tumor response was estimated using objective tumor measurements which are measured using imaging methods and determined according to the Response Evaluation Criteria in Solid Tumors guidelines \cite{therasse2000Tumors} or the World Health Organization recommendations \cite{world1979handbook}. 
	The RCTs in the systematic review contain five treatment classes: the class of chemotherapies, the anti-epidermal growth factor receptor (Anti-EGFR) monoclonal antibodies class, angiogenesis inhibitors, other molecular-targeted agents (MTA) and intrahepatic arterial (IHA) chemotherapies .
	
	Ciani et al. \cite{ciani2015meta} investigated surrogate relationships between treatment effects on potential surrogate endpoints (TR and PFS) and on the final clinical outcome (OS). They found that the surrogate relationships between treatment effects on these endpoints were suboptimal. Furthermore, they stated that PFS was an acceptable surrogate endpoint for OS, whereas TR should not be used as a surrogate endpoint for this final outcome. They concluded that good surrogacy observed in previous studies, that included traditional chemotherapy trials in aCRC may not apply directly across other classes of treatments. More details about the studies and how the systematic review was designed can be found in Ciani et al. \cite{ciani2015meta}. We refer these data as 'Ciani data' in the remainder of this paper.
	
	In our example, we focused on a subset of these data examining the surrogacy between treatment effects on TR and PFS and treatment effects on PFS and OS including data from three treatment classes. We obtained data from 35 studies reporting treatment effect on PFS and OS where, 15 of them belonged to the chemotherapy treatment class, 9 of them investigated anti-EGFR therapies and 11 anti-angiogenic treatments. To investigate surrogate relationships between treatment effects on TR and PFS we used data from 35 studies reporting treatment effects on these endpoints; 17 of them investigated chemotherapies, 8 and 10 studies anti-EGFR and anti-angiogenic treatments respectively. TR can be evaluated as a surrogate endpoint to treatment effect on PFS, as treatment effects on TR is typically measured earlier compared to treatment effects on PFS. 
	
\begin{figure}[h!]
		\caption{Scatterplots of treatment effects on PFS-OS and TR-PFS}
		\label{fig:1}
		\centering
		\includegraphics[height=7cm,width=15cm]{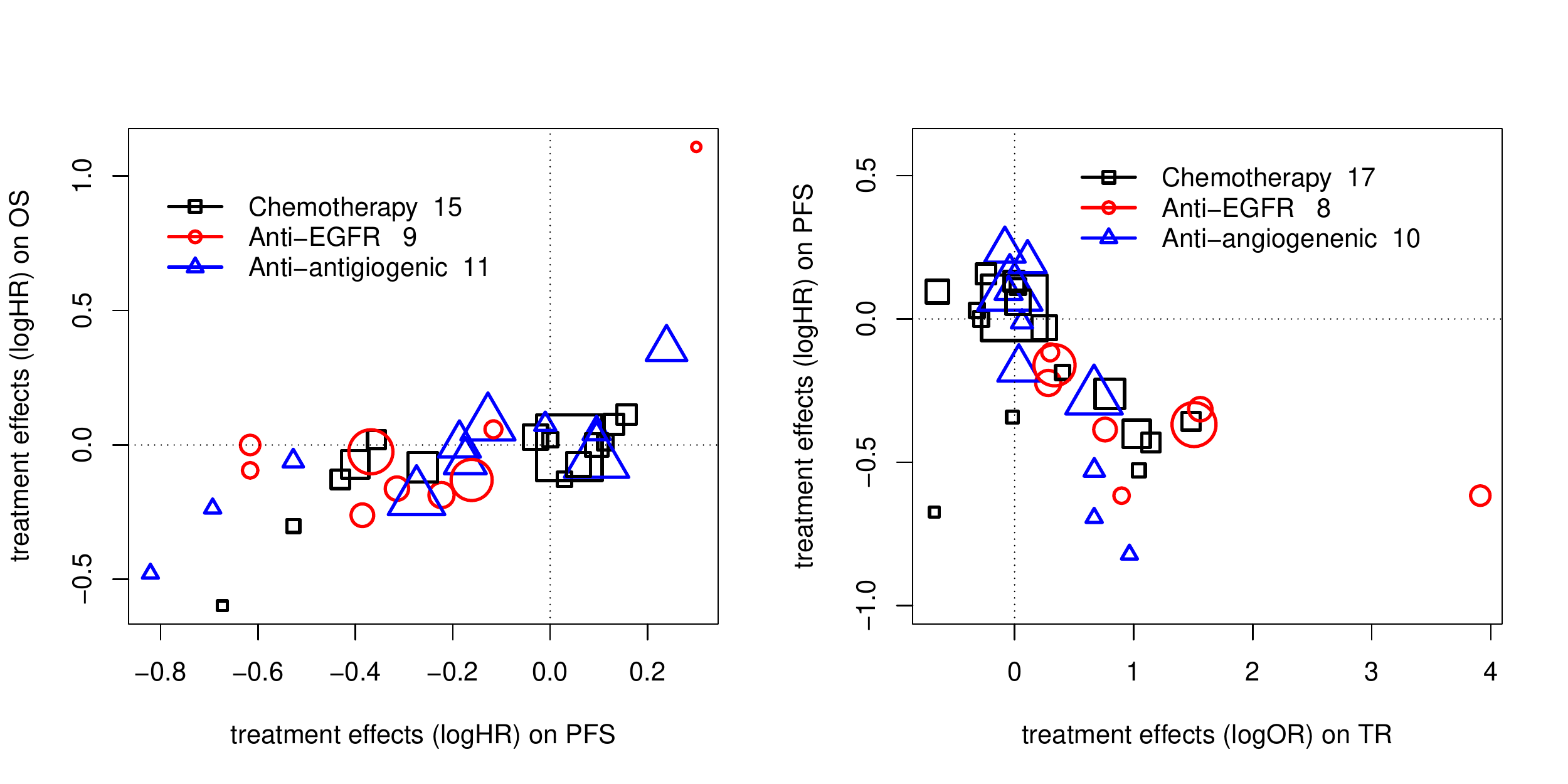}
	\end{figure}

	Figure \ref{fig:1} provides a graphical representation of the data set we used. It illustrates the association patterns between the treatment effects across classes on each pair of outcomes.
	
	Individual patient data (IPD) were available from four RCTs\cite{bennouna2013continuation,rothenberg2008capecitabine,cassidy2011xelox,hurwitz2004bevacizumab}, which were used to estimate the within-study correlations. By applying a bootstrap method (see section A in the supplementary material) we estimated two sets of within-study correlations: for each of the two pairs of outcomes one correlation corresponding to each treatment class. We assumed that within treatment classes the within-study correlations are the same across studies.

	\subsection{Scale of the outcomes}
	The treatment effects on OS and PFS were modelled on the log hazard ratio scale $logHR(OS)$, $logHR(PFS)$, whereas the treatment effects on TR were modelled on log odds ratio $logOR(TR)$ scale. We retrieved the corresponding standard errors of $logHR(PFS)$ and $logHR(OS)$ on PFS and OS from the 95\% confidence intervals and by using the standard formulae for the standard errors of $logOR(TR)$ .

	\subsection{Results of data analysis}\label{ApplicResults}
	The first aim of our analysis was to explore potential differences in association patterns across treatment classes. To investigate this, we applied the two proposed models and subgroup analysis using standard model to the data and derived posterior distributions for the parameters of the surrogate relationships for each treatment class. We obtained the posteriors mean of the intercepts $\hat{\lambda}_{0j}$, the slopes $\hat{\lambda}_{1j}$ and posterior median of conditional variances $\hat{\psi}^{2}_{j}$ with corresponding 95\% CrIs across treatment classes. By checking the surrogacy criteria (described in section \ref{Criteria}) we were able to infer whether or not a candidate endpoint is a valid surrogate in each treatment class. We carried out a cross-validation procedure (section \ref{CrossVal}) to investigate how well the models predict the true treatment effect on the final clinical outcome. The measures we monitored were the absolute error of the predictions, the ratios of the width of the 95\% predictive intervals from P-EX or F-EX to the width of the 95\% predicted interval obtained from subgroup analysis and the largest MCE. 
	\newpage
	\subsubsection{Results across models and treatment classes}
	\ \newline
	\textbf{Subgroup analysis with the standard model}
	
	The results of subgroup analysis presented in the first two columns of Table \ref{Estimates} showed strong association between the treatment effects on PFS and the effects on OS in the class of chemotherapies and the anti-angiogenic treatment class with all three criteria for surrogacy satisfied (the 95\% CrIs of $\lambda_{01}$ and $\lambda_{03}$ included zero, the 95\% CrIs of $\lambda_{11}$ and $\lambda_{13}$ did not contain zero and there was substantial evidence using Bayes factors in favour of the hypotheses $H_{1}:\psi_{1}^{2}=0$, and $H_{1}:\psi_{3}^{2}=0$ (see details about Bayes factors in the supplementary material D.2)). In contrast, we can infer that the surrogate relationship between treatment effects on PFS and the effects on OS in the anti-EGFR treatment class was weak, as the 95\% CrI of the posterior distribution of the slope included zero. 
	Investigating the surrogacy on TR-PFS pair we found a similar pattern, thus we can infer that there was an acceptable surrogate relationship between treatment effects on TR and PFS in the chemotherapy and the anti-angiogenic classes. The relationship was negative overall, since the slopes were negative across classes. On the other hand, the surrogacy criteria indicated poor surrogacy between the treatment effects on TR and the treatment effects on PFS for anti-EGFR class, since the 95\% CrI  of the slope $\lambda_{12}$ included zero.
	\newline
	\ \newline
\textbf{F-EX model}
	
	The results of F-EX model are presented in columns 3 and 4 of Table \ref{Estimates}. For the PFS-OS pair of outcomes, the association patterns were very similar in the anti-angiogenic and chemotherapy treatment classes as both classes satisfied the surrogacy criteria and the slopes were of similar magnitude. The 95\% CrIs of the intercepts $\lambda_{01}$ and $\lambda_{03}$ included zero indicating that zero treatment effect on the surrogate implies zero treatment effect on the final outcome for these two classes. The intervals of the slopes $\lambda_{11}$ and $\lambda_{13}$ did not contain zero indicating positive association as the two slopes were positive. The conditional variances in these two classes were small indicating strong association which was supported by the analysis using Bayes factors (see details about the Bayes factors in the supplementary material D.2). On the other hand, the association was weak in the anti-EGFR treatment class failing to meet one of the criteria, as the 95\% CrI of the slope $\lambda_{12}$ included zero.
	On the contrary, for TR-PFS pair of outcomes all three surrogacy criteria were satisfied across all the treatment classes taking advantage of the assumption of exchangeability for the parameters $\lambda_{0j}$ and $\lambda_{1j}$. This implies that TR was an acceptable surrogate endpoint for PFS across treatment classes in this data set.
	\newline
	\ \newline
	\textbf{P-EX model}
	
	P-EX model allows the parameters of slope of each treatment class to be either exchangeable or non-exchangeable with parameters of slopes from other classes yielding parameters with partial exchangeability. For both pairs of outcomes, fixed values for the hyper-parameters $\pi_{j}=(0.5,0.5,0.5)$ were chosen assuming that exchangeability and non-exchangeability were $\textit{apriori}$ equally likely.
	
	As in the case of F-EX model, the surrogacy criteria were estimated for each class separately and then a cross-validation procedure followed, however, for this model we also monitored the mixture weights by calculating the posterior means of $p_{j}$ in order to measure the degree of borrowing of information across classes (Table \ref{Estimates} columns 5, 6). For the PFS-OS pair, the weights increased from their prior values ($\pi_{j}=0.5$) to 0.968 in the class of chemotherapy, to 0.965 in the anti-EGFR class and to 0.966 in the anti-angiogenic treatment class indicating that borrowing of information was reduced approximately 3.5\% for each class compared to F-EX model. Looking at the results from P-EX model we drew the same inferences as from F-EX model, inferring that the association patterns were strong in the anti-angiogenic and the chemotherapy classes, but weak in the anti-EGFR treatment class where the 95\% CrI of the slope $\lambda_{12}$ included zero. 
	In contrast to this, for TR-PFS pair the mixture weights were smaller than on PFS-OS pair due to the slightly larger between treatment class heterogeneity. There was 7.1\% reduction in borrowing of information in anti-angiogenic class compared to F-EX models, whilst the weights for the chemotherapies and anti-EGFR agents were 0.944 and 0.95 respectively. All three surrogacy criteria were fulfilled across treatment classes despite the decrease in levels of borrowing of information, indicating that TR was an acceptable surrogate for PFS across treatment classes in the Ciani data.
	
	\begin{table}[h!]	
		\fontsize{9}{16}\selectfont
			\caption{Estimates of the parameters defining the surrogacy criteria}
			\label{Estimates}
			\centering
			\resizebox{\linewidth}{!}{
			\begin{tabular}{lcccccc}
				&\multicolumn{2}{c}{$\textbf{Standard model}$}&\multicolumn{2}{c}{$\textbf{F-EX}$}&\multicolumn{2}{c}{$\textbf{P-EX}$}\\
				\hline
				\shortstack{Treatment \\ Classes}    &PFS-OS & TR-PFS &PFS-OS & TR-PFS & PFS-OS & TR-PFS  \\
				\hline
				$\textbf{\textit{chemotherapy}}$     & N=15& N=17& N=15& N=17& N=15& N=17\\
					$p_{1}$  &-&-&-&-& 0.968&0.944\\
				$\lambda_{01}$    & -0.002 (-0.059,{ }0.053)          & -0.05 (-0.164,{ }0.033)     & 0.003 (-0.050,{ }0.054) & -0.051 (-0.154,{ }0.033)          & 0.003 (-0.050,{ }0.054)   & -0.051 (-0.155,{ }0.033)                  \\
				$\lambda_{11}$  & { }0.322 ({ }0.089,{ }0.548)      &-0.261 (-0.402,-0.097)             & 0.334 ({ }0.124,{ }0.533)         &-0.267 (-0.406,-0.111)             & 0.334 ({ }0.124,{ }0.535) &-0.266 (-0.404,-0.109)                     \\
				$\psi^{2}_{1}$  & 0.001 ({ }5$\cdot10^{-6}$,0.009) &{ }0.016 ({ }4$\cdot10^{-4}$,0.072)& 0.001 ({ }2$\cdot10^{-6}$,0.009) &{ }0.016 ({ }4$\cdot10^{-4}$,0.069)& 0.0007 ({ }2$\cdot10^{-6}$,0.009) &{ }0.016 ({ }3$\cdot10^{-4}$,0.069)\\
				$\textbf{\textit{anti-EGFR}}$  &N=9&N=8&N=9&N=8&N=9&N=8     \\
				$p_{2}$  &-&-&-&-&0.965&0.950\\
				$\lambda_{02}$  & -0.048 (-0.292,{ }0.296)   &  -0.195 (-0.415,{ }0.033)                 & 0.001 (-0.153,{ }0.146)   &  -0.138 (-0.338,{ }0.059)                  & -0.001 (-0.160,{ }0.149) &  -0.140 (-0.341,{ }0.058)                          \\
				$\lambda_{12}$  &{ }0.126 (-0.544,{ }1.031)&  -0.140 (-0.366,{ }0.019)                   &0.274 (-0.157,{ }0.640)&  -0.187 (-0.421,-0.027)                        &{ }0.268 (-0.182,{ }0.648)&  -0.184 (-0.418,-0.026)                            \\
				$\psi^{2}_{2}$  & 0.008 ({ }2$\cdot10^{-5}$,0.103)  & { }0.013 ({ }7$\cdot10^{-5}$,0.131)& 0.010 ({ }8$\cdot10^{-5}$,0.078)  & { }0.014 ({ }8$\cdot10^{-5}$,0.128)& { }0.010 ({ }8$\cdot10^{-5}$,0.079)  & { }0.014 ({ }7$\cdot10^{-5}$,0.127)\\
				$\textbf{\textit{anti-angiogenic}}$&N=11&N=10&N=11&N=10&N=11&N=10 \\
				$p_{3}$  &-&-&-&-&0.966&0.929\\
				$\lambda_{03}$  &0.052 (-0.038,{ }0.149)  &   { }0.074 (-0.079,{ }0.225)                &0.031 (-0.041,{ }0.113)  &   { }0.030 (-0.131,{ }0.178)                  &0.032 (-0.041,{ }0.115)  &   { }0.031 (-0.132,{ }0.180)                  \\
				$\lambda_{13}$  &0.481 ({ }0.174,{ }0.797)  & -0.786 (-1.197,-0.455)                    &0.411 ({ }0.158,{ }0.685)  & -0.674 (-1.060,-0.271)                      &0.413 ({ }0.160,{ }0.694)  & -0.686 (-1.075,-0.280)                      \\
				$\psi^{2}_{3}$  & 0.006 ({ }1$\cdot10^{-4}$,0.040)  &{ }0.011 ({ }4$\cdot10^{-5}$,0.092)&0.006 ({ }6$\cdot10^{-5}$,0.036)  &{ }0.015 ({ }1$\cdot10^{-4}$,0.115)   &0.006 ({ }8$\cdot10^{-5}$,0.036)  &{ }0.015 ({ }6$\cdot10^{-5}$,0.114)\\
				\hline
				
			\end{tabular}
		}
	\end{table}	
\newpage
	\subsubsection{Results of the cross-validation procedure}
	After estimating the surrogacy criteria across treatment classes, we carried out cross-validation procedure to predict the treatment effects $\mu_{2i}$ on the final outcome. The results in Table \ref{CV} showed that the cross-validation procedure of subgroup analysis with the standard model gave predictive intervals of the effects on the final outcome containing the corresponding observed estimates $Y_{2i}$ in the 97\% of the studies for both pairs of outcomes confirming good fit of the model. The cross-validation procedure yielded the most accurate posterior means of the true effects on the final endpoint (small absolute error) in the treatment class of chemotherapies, where the number of the available studies was large and performed poorly in terms of accuracy of predictions in the anti-EGFR class (large absolute error) where the surrogacy was weak and the number of studies small.
	Similarly, subgroup analysis with the standard model was less accurate in targeted treatment classes for the TR-PFS pair of outcomes where the number of studies was smaller. 
	
	The results from the cross-validation procedure of F-EX model showed that the method fitted the data well. All of the predicted intervals of $\mu_{2ij}$ contained the observed values of the treatment effects on the final outcome on PFS-OS pair and all but one on TR-PFS pair. The cross-validation procedure yielded the posterior means of $\mu_{2ij}$ with the smallest absolute error in chemotherapy treatment class on PFS-OS pair and performed equally well in terms of its accuracy in the other two classes. In contrast to this, higher absolute error were observed in the anti-angiogenic class on TR-PFS pair indicating that the assumption of exchangeability for the parameters describing the surrogate relationships was fairly strong and it was likely to cause 'overshrinkage' in this particular class.
	The results obtained for the width ratios imply that F-EX method gave intervals of the true effect on the final endpoint with smaller degree of uncertainty compared to subgroup analysis.
	There was a small decrease in the uncertainty of the predictions of $\mu_{2ij}$ on PFS-OS pair for the chemotherapy treatment class, as the cross-validation procedure of F-EX model yielded 1.2\% narrower intervals compared to subgroup analysis. Furthermore, significantly reduced uncertainty was observed in the other two treatment classes for PFS-OS pair, 13.8\% in the anti-EGFR treatment class and 7\% in the anti-angiogenic, where the number of studies was smaller.
	On the contrary, very limited decrease in the degree of uncertainty was observed for the TR-PFS pair of outcomes across all classes. Overall on this pair, the predictive intervals were only 1.3\% narrower compared to subgroup analysis. The benefit was small (3.2\% reduction of the width of the predictive interval) even for the anti-EGFR treatment class where there were only 8 studies for this pair.

	Focusing on the results from the cross-validation procedure using P-EX model, all the intervals of the predicted treatment effects on the final outcome contained the observed treatment effects on PFS-OS pair and all but one on the TR-PFS pair. The absolute error was smaller in chemotherapy treatment class for the PFS-OS pair where the number of studies was large and significantly higher in the other two classes. In contrast to this, the cross-validation procedure with P-EX gave almost equally accurate estimates in the anti-EGFR and the chemotherapy treatment classes on TR-PFS pair. However, the absolute error was higher in the anti-angiogenic treatment class where the association was much stronger compared to the other two classes indicating potential excessive borrowing of information from the other classes. This is likely due to the assumption of full exchangeability for the intercepts. 
	
	The method predicted the effects on the final outcome with reduced uncertainty giving more precise estimates ($\hat{\mu}_{2ij}$) compared to subgroup analysis in the anti-EGFR class on PFS-OS pair reducing the uncertainty by 13.6\%. On the other hand, the predicted effects $\hat\mu_{2ij}$ had almost the same degree of uncertainty as those from subgroup analysis for TR-PFS pair. The intervals were only 1\% narrower on average across all classes compared to the subgroup analysis. 
	
	\begin{table}[h!]
		\centering
		\fontsize{9}{15}\selectfont
		\caption{Predictions of $\mu_{2ij}$ across treatments and models}
		\label{CV}
		\begin{tabular}{llcccccccccc}
			\hline
			&&\multicolumn{2}{c}{$\textbf{\textit{chemotherapy}}$}&\multicolumn{2}{c}{$\textbf{\textit{anti-EGFR}}$}&\multicolumn{2}{c}{$\textbf{\textit{anti-angiogenic}}$}&&\multicolumn{2}{c}{$\textbf{\textit{Overall}}$}&\\
			\cline{3-4}\cline{5-6}\cline{7-8}\cline{10-11}
			Models&Measures   &PFS-OS& TR-PFS&PFS-OS& TR-PFS&PFS-OS& TR-PFS&&PFS-OS&TR-PFS&\\
			\hline
			\shortstack{Standard\\ Model }&\shortstack{\\Performance of  95\% \\predictive intervals \ \ }  & 1.000&0.941&0.888&1.000&1.000&1.000&&0.971&0.971&\\
			&Absolute error (median) &0.047   &0.108&0.140&0.132&0.099&0.145&&0.090&0.123&\\
			&MCE (max) & 0.002 & 0.004&0.006&0.004&0.003&0.004&&0.003&0.004& \\
			\hline
			\shortstack{F-EX\\ \ }&\shortstack{\\Performance of  95\% \\predictive intervals \ \ }  & 1.000&0.941&1.000&1.000&1.000&1.000&&1.000&0.971&\\
			&Absolute error (median) & 0.041  &0.104&0.102&0.112&0.123&0.206&&0.089&0.128&\\
			&Width ratio (median) &0.988&0.985&0.862&0.968&0.930&0.997&&0.950&0.987&\\
			&MCE (max) & 0.002 & 0.004&0.005&0.005&0.003&0.003&&0.003&0.004& \\
			\hline
			\shortstack{P-EX\\ \ }&\shortstack{\\Performance of  95\% \\predictive intervals \ \ }  & 1.000&0.941&1.000&1.000&1.000&1.000&&1.000&0.971&\\
			&Absolute error (median)& 0.041  &0.104&0.126&0.114&0.109&0.206&&0.092&0.128&\\
			&Width ratio (median)&0.989&0.989&0.864&0.975&0.931&0.999&&0.957&0.990&\\
			&MCE (max) & 0.002 & 0.004&0.005&0.005&0.003&0.003&&0.003&0.004& \\	
			\hline	
		\end{tabular}	
	\end{table}
		
	\subsection{Comparison of the results from F-EX, P-EX and those from subgroup analysis}
	Figure 2 presents 95\% CrIs of the slopes $\lambda_{1j}$ and intercepts $\lambda_{0j}$ across the treatment classes and methods of estimation. Comparing the aforementioned methods in regards to the surrogacy criteria on the PFS-OS pair, we can conclude that F-EX model estimated the parameters of the surrogate relationships with reduced uncertainty compared to the subgroup analysis and P-EX model taking advantage of borrowing of information across classes. P-EX relaxes the assumption of exchangeability reducing the effect of borrowing of information on average by 3.6\%. It gave narrower CrIs of the parameters of interest compared to subgroup analysis but slightly larger than those obtained form F-EX model. Furthermore, both F-EX and P-EX methods can distinguish between the different association patterns avoiding to give over-shrunk estimates of the slopes and the intercepts, although they allow different degrees of borrowing of information for the slopes. In particular, this pair of outcomes (PFS-OS) illustrates well the impact of number of studies per class on the degree of borrowing of information. In general, borrowing of information is determined by the number of studies within treatment classes, between treatment classes heterogeneity, as well as the number of treatment classes. In this case, the fewer studies we have within a treatment class, the bigger is the impact of borrowing of information resulting in higher reduction in uncertainty of the estimates of surrogate relationships. This effect was particularly strong for the anti-EGFR treatment class.

	\begin{figure}[h!]
			\centering
			\caption{95\% Credible intervals of $\lambda_{1j}$ and $\lambda_{0j}$ for the PFS-OS pair of outcomes}
			\label{fig:3}
			\includegraphics[height=8.5cm, width=16.57cm]{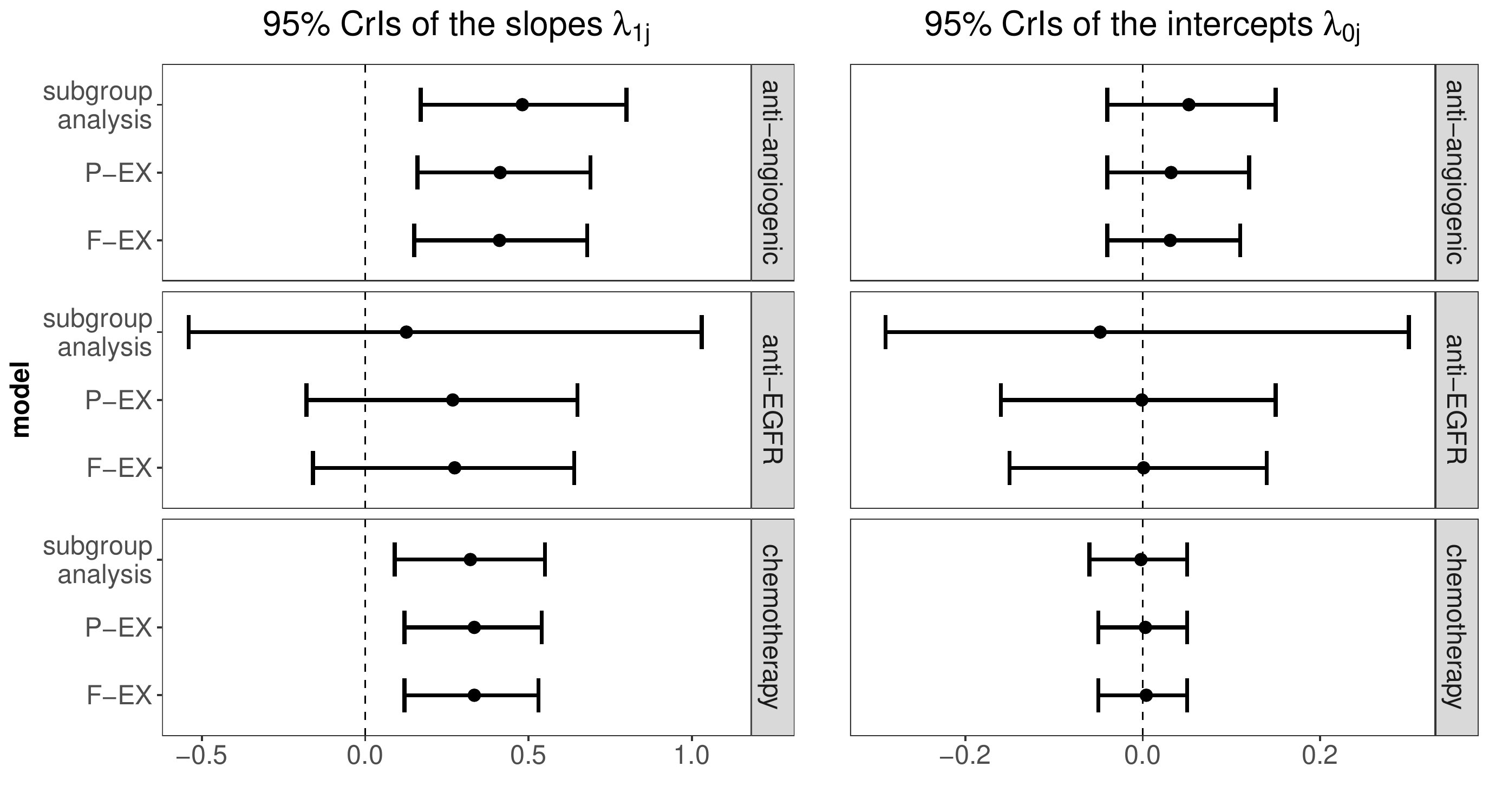}
		
	\end{figure}

	On the other hand, TR-PFS pair is a good example to illustrate the performance of the hierarchical methods when between treatment class heterogeneity is relatively large. In this case, subgroup analysis performed equally well as the proposed methods in terms of uncertainty of the CrIs of the paramaters describing the surrogate relationships. For instance by fitting F-EX and P-EX models, we did not observe any decrease in uncertainty around $\lambda_{1j}$ and $\lambda_{0j}$ across classes. This is because the between treatment classes heterogeneity was relatively large for TR-PFS pair and hence there was not much shrinkage. Furthermore using subgroup analysis, the surrogacy criteria failed in the anti-EGFR class (zero was included in the 95\% CrI of the slope) where only 8 studies available). However, the 95\% CrI in the anti-EGFR class just contains zero and overlaps substantially with the 95\% CrI of the slope for chemotherapy treatment class.
	By applying P-EX and F-EX models, we were able to draw different inferences for the surrogacy in the anti-EGFR class as these methods allow for borrowing of information for the parameters describing the surrogate relationships from the other classes. As illustrated in Figure 3, both hierarchical models moved the 95\% CrI of the slope in the direction of the CrIs of the other two classes resulting in the surrogacy criteria being satisfied across all treatment classes.

	\begin{figure}[h!]
			\centering
			\caption{95\% Credible intervals of $\lambda_{1j}$ and $\lambda_{0j}$ for the TR-PFS pair of outcomes}
			\label{fig:4}
			\includegraphics[height=8.5cm, width=16.57cm]{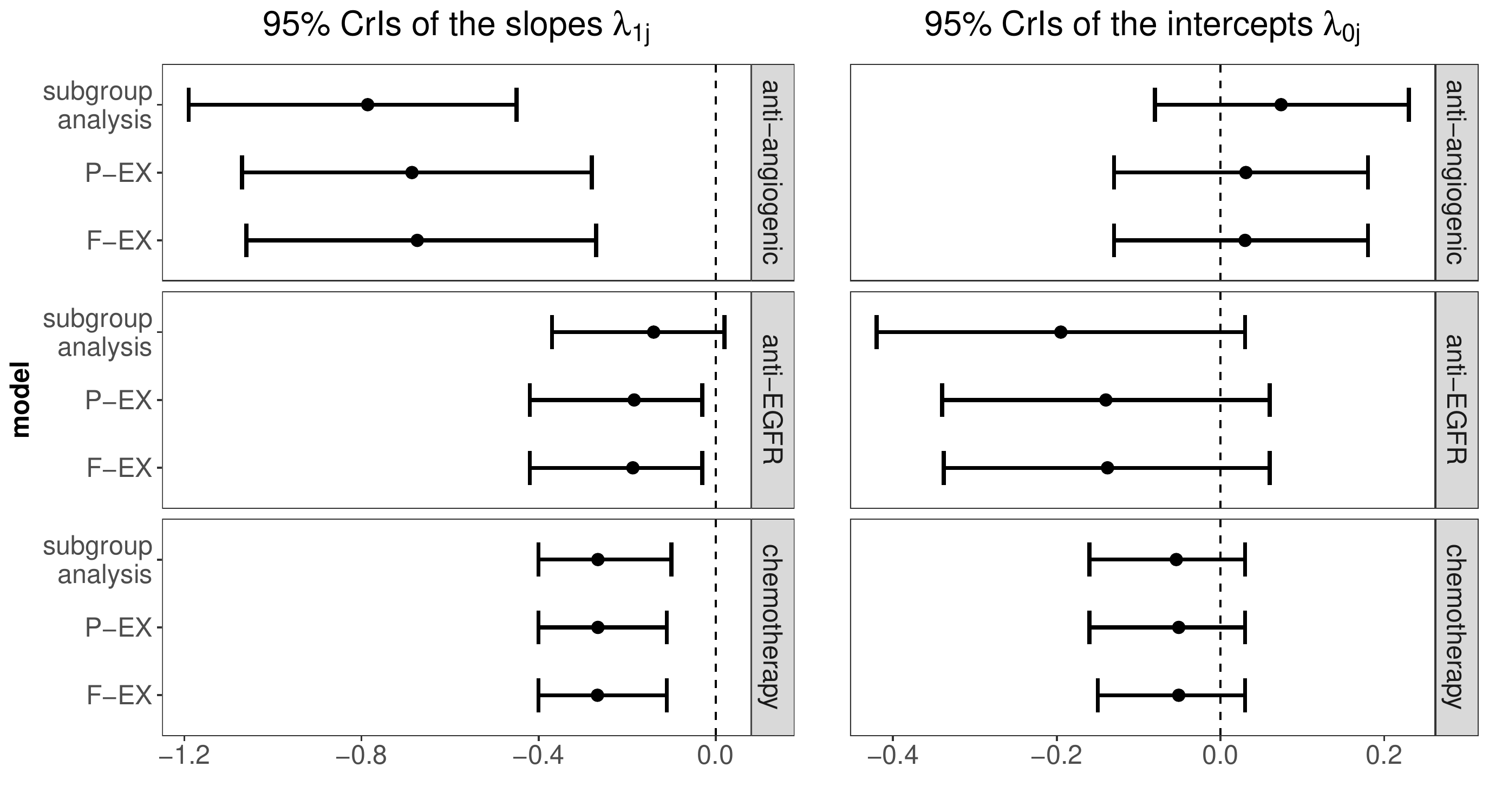}
	\end{figure}

	When carrying out cross-validation procedure, we wish to ensure that not only predictive intervals contain the observed values but also that they are sufficiently narrow. In general, adding a hierarchical structure to slopes and intercepts reduces the uncertainty and leads to more precise predictions compared to those obtained from subgroup analysis.
	For the PFS-OS pair of outcomes, the accuracy of the predictions was very similar across all methods (similar absolute error) but the uncertainty varied depending of the level of borrowing of information. F-EX model gave on average the most precise estimates ($\hat{\mu}_{2ij}$) having the narrowest 95\% predictive intervals of the effect on the final outcome (smallest width ratio seen in Tables 3, 5, 7) reducing the overall uncertainly by 5\%. The benefit was smaller in the chemotherapy class where the number of studies was much larger compared to the anti-EGFR treatment class where we had only 8 studies available. Overall, P-EX performed better than subgroup analysis and equally well with F-EX regarding the uncertainty of the predictions. This indicates that the assumption of exchangeability seems to be plausible for this pair of outcomes and P-EX model was able to identify this.
	
	For the TR-PFS pair, subgroup analysis with the standard model was a robust approach in terms of the accuracy of its predictions. Although the overall absolute error was very similar across models, F-EX and P-EX yielded higher absolute error compared to subgroup analysis in the anti-angiogenic class. This implies that the posterior means of the true effects were to some extent 'overshrunk' due to excessive borrowing of information from the other classes. P-EX model was implemented allowing for partial exchangeability only for the slopes, this decision is likely to affect the performance of the model in terms of its predictions on TR-PFS pair of outcomes. However, the model can be extended allowing for partial exchangeability also of the intercepts or the conditional variances and different combinations of these assumptions can be explored and models compared using DIC. 
	Similarly, there was no significant decrease in the degree of uncertainty of the estimates $\hat{\mu}_{2ij}$ of F-EX and P-EX models.
	The results indicate that the hierarchical methods performed slightly better compared to subgroup analysis in terms of uncertainty only in the class of chemotherapy and the anti-EGFR treatment class giving 1.5\% and 3\% narrower predictive intervals respectively. This kind of behaviour might be caused by the relatively large between treatment class heterogeneity and the assumption of full exchangeability of the intercepts.

	\section{Discussion}
	We developed two hierarchical models allowing to account for distinct treatment classes when examining the surrogate relationships. The proposed models may be particularly useful in surrogate endpoint evaluation in complex diseases where different treatment classes of different mechanism of action and potential different association patterns within those classes exist. These models investigate potential differences in study level surrogacy across treatment classes in a particular disease area and can help to identify treatment classes with strong association patters, even when data are relatively sparse. F-EX model is somewhat restrictive, assuming full exchangeability for the parameters describing the surrogate relationships across treatment classes. In many situations the assumption of exchangeability may be too strong given the heterogeneity between treatment classes. In such circumstances, a more flexible model such as P-EX may be a better choice. P-EX model can infer an appropriate level of borrowing of information from the data, reducing the degree of borrowing of information through the mixture weights, thus relaxing the assumption of exchangeability when it is not fully reasonable. It evaluates whether the association pattern between treatment effects (logHR or logOR) on the surrogate and the final endpoint in a specific treatment class differs from the other patterns in other classes. 
	
	F-EX model is appropriate only when the degree of similarity of surrogate relationships is relatively high. It can offer substantial gains in precision, reduced RMSE of the posterior means of the parameters describing surrogate relationships and it can improve the predictions of the true effects on the final endpoint. For example, F-EX model gave posterior means of the slopes and predicted effects with reduced uncertainty (smaller credible intervals) compared to subgroup analysis for the first simulated data scenario and for the illustrative example on PFS-OS pair where the parameters describing the surrogate relationship were similar and the assumption of full exchangeability was reasonable. These findings are consistent with the results from other hierarchical Bayesian methods which assume full exchangeability and were developed in other research areas \cite{berry2013Hierarchical,thall2003}. However, P-EX model achieves the same degree of borrowing of information in such data scenarios making less assumptions compared to F-EX model. P-EX model regulates the degree of borrowing of information using its exchangeable and non-exchangeable components with respective mixture weights. For instance, when between treatment class heterogeneity is relatively large or there is a treatment class with distinctly different pattern, P-EX model has the advantage of avoiding the excessive borrowing of information, as illustrated in the second design of the simulation study.  All the above illustrate the benefits of partial exchangeability, as  described by Neuenschwander et al. \cite{neuenschwander2016robust} in their work. 
	Subgroup analysis using the standard model is a simple method which performs well when there are sufficient data available for each treatment class, but it produces estimates with higher bias and uncertainty when data within a treatment class are limited. In any other situation P-EX should be preferred as it regulates the degree of borrowing of the slopes.
	
	Although the proposed methods provide additional robustness to the CrIs and the posterior means of the parameters describing the surrogate relationships compared to subgroup analysis, potential limitations should always be kept in mind. First, in real data scenarios it can be challenging to find data sets with sufficient number of treatment classes. The small number of treatment classes can affect the performance of hierarchical methods substantially \cite{mcneish2016modeling} reducing the impact of borrowing of information. For instance, fitting P-EX model to the illustrative example (in aCRC with three treatment classes) led to a situation where in some of the MCMC iterations only one class was deemed exchangeable by the model which is not possible since there were no other classes to exchange information with. However, in our example it did not affect the performance of the model as it occurred only in the 0.5\% of the MCMC iterations. On the other hand, there is no upper limit to the number of classes we can have. In general, the more classes the better it is for the models to borrow information across them. 
	
	Another limitation of the illustrative example is that treatment switching was applied in a subset of trials in this data set. Patients were allowed to switch from the treatment that was initially assigned to them to the other treatment arm in the trial. Most commonly patients switched after progression from control to experimental arm in particular, if there was sufficient evidence during the trial that the experimental treatment was better than control \cite{latimer2016treatment}. Treatment switching has diminishing effect on the difference in treatment effects on OS when applying intention-to-treat analysis, and the effect is often obtained with larger uncertainty. This makes the estimation of surrogacy between treatment effects on the surrogate and treatment effects on the final outcome very challenging. Many adjustment methods have been proposed, however, their validity is often questionable\cite{latimer2016treatment}. Additionally, the evaluation of PFS as a surrogate endpoint is distinctive compared to other surrogate endpoints as PFS can be considered as nested outcome within OS outcome. These factors may explain the different findings for the two pairs of outcomes (PFS-OS and TR-OS).
	
	Furthermore, as it was mentioned in section \ref{Application}, each treatment class consist of studies with multiple treatment comparisons. According to Daniels and Hughes \cite{daniels1997meta} and Shanafelt et. al  \cite{shanafelt2004chemotherapy} different treatment comparisons and the use of active or inactive control interventions may influence the surrogate relationship. This could potentially be resolved by classifying treatment according the treatment class comparison (for example anti-angiogenic therapies versus chemotherapy) which potentially would lead to more treatment classes, but with reduced number of studies per class. To continue with the same issue, in this paper the treatment classes were defined according to the class of the experimental treatment regardless the control. Alternatively, we could classify them according to the treatment contrasts taking into account the class of the control group, however, this could result in fewer studies per class. A network meta-analysis model was developed for this problem by Bujkiewicz et al. \cite{bujkiewicz2019bivariate}. 
	
	Additionally the evaluation framework proposed by Daniels and Hughes (see section \ref{Criteria}) examine whether zero is contained in the CrIs of $\lambda_{1}$ and $\lambda_{0}$. However, the sparsity of data may lead to increased uncertainty around the intercept and slope. This increased uncertainty is also likely to manifest itself in increased conditional variance, thus invalidating the third criterion. Unsurprisingly, for sparse data it is unlikely that all the surrogacy criteria hold and this problem is more likely to occur in subgroup analyses. Our proposed methods alleviate this problem as shown in some of the scenarios of the simulation study. However, we used the criteria mainly for the purpose of model comparison. In real life scenarios, when evaluating a potential surrogate endpoint for use in clinical trials or regulatory decision making, the decision of whether the surrogate endpoint may be used to make the prediction of the clinical benefit should be based on the balance between the strength of the surrogate relationship and the need for the decision to be made about the effectiveness of the new treatment \cite{alonso2016applied}. Moreover, the strength (or weakness) of the surrogate relationship will manifest itself in the width of the predicted interval of the treatment effect on the final outcome. A larger interval around the intercept and slope will result in a larger interval around the predicted effect and hence increased uncertainty about the regulatory or clinical decision made based on such prediction. The implication of this is that perhaps we do not need precise surrogacy criteria and instead we need only look at the predictions \cite{bujkiewicz2019bivariate}. The quality of predictions can be evaluated through a cross-validation procedure (see section \ref{CrossVal}).

	A possible extension of these methods is to add another layer of hierarchy accounting for the different treatments within a treatment class. However, a relatively large number of studies for each treatment and number of treatments per class would be required to fit such model. As we mentioned in section 6.4, P-EX model could also be extended by making additional partial-exchangeability assumptions about the intercepts and the conditional variances, however, this may lead to over-parameterising the model.
	Furthermore, taking advantage of the setting proposed by Bujkiewicz et al. \cite{bujkiewicz2015bayesian}, both hierarchical models can be extended to allow for modelling  multiple surrogate endpoints (or the same surrogate endpoint but reported at multiple time points) as joint predictors of treatment effect on the final outcome.
		
	Further research is also needed to extend the proposed methodology to Binomial data or to time to event data where the assumption of normality is not plausible.
	Moreover, to overcome the convergence issues caused by vague prior distributions on the hyper-parameter of the mixture weights ($\pi_{j}$), alternative prior distributions should be developed by extending the P-EX in a similar way as proposed by Kaizer et al. \cite{kaizer2017bayesian}.
	
	In summary,	we developed hierarchical Bayesian methods for evaluating surrogate relationships within treatment classes whilst borrowing of information for surrogate relationships across treatment classes. We believe that the proposed methods have a lot of potential for improving the validation of surrogate endpoints in the era of personalized medicine, where the surrogacy may depend on the mechanism of action of specific targeted therapies.

	\section{Acknowledgements}
	
This research used the ALICE/SPECTRE High Performance Computing Facility at the University of Leicester and was partly funded by the Medical Research Council, grant no.  MR/L009854/1 awarded to Sylwia Bujkiewicz. Keith Abrams is partially supported as a UK National Institute for Health Research (NIHR) Senior Investigator Emeritus (NI-SI-0512-10159). We also thank the anonymous reviewers for their comments, which helped to improve the quality of the manuscript.

\bibliography{bibli}

\begin{thebibliography}{10}
\providecommand \doibase [0]{http://dx.doi.org/}%

\bibitem{burzykowski2006evaluation}
Burzykowski T, Molenberghs G, Buyse M. {\it The evaluation of surrogate
  endpoints}.
\newblock Springer Science \& Business Media .
\newblock 2006.

\bibitem{fleming2012biomarkers}
Fleming TR, Powers JH. Biomarkers and surrogate endpoints in clinical trials.
  {\it Statistics in medicine} 2012\string; 31(25)\string: 2973--2984.

\bibitem{phillips2003e9}
Phillips A, Haudiquet V. ICH E9 guideline Statistical principles for clinical
  trials: a case study. {\it Statistics in medicine} 2003\string; 22(1)\string:
  1--11.

\bibitem{daniels1997meta}
Daniels MJ, Hughes MD. Meta-analysis for the evaluation of potential surrogate
  markers. {\it Statistics in medicine} 1997\string; 16(17)\string: 1965--1982.

\bibitem{buyse2000validation}
Buyse M, Molenberghs G, Burzykowski T, Renard D, Geys H. The validation of
  surrogate endpoints in meta-analyses of randomized experiments. {\it
  Biostatistics} 2000\string; 1(1)\string: 49--67.

\bibitem{bujkiewicz2015uncertainty}
Bujkiewicz S, Thompson JR, Spata E, Abrams KR. Uncertainty in the Bayesian
  meta-analysis of normally distributed surrogate endpoints. {\it Statistical
  methods in medical research} 2015\string; 26(5)\string: 2287--2318.

\bibitem{lassere2008biomarker}
Lassere MN. The Biomarker-Surrogacy Evaluation Schema: a review of the
  biomarker-surrogate literature and a proposal for a criterion-based,
  quantitative, multidimensional hierarchical levels of evidence schema for
  evaluating the status of biomarkers as surrogate endpoints. {\it Statistical
  methods in medical research} 2008\string; 17(3)\string: 303--340.

\bibitem{buyse2007progression}
Buyse M, Burzykowski T, Carroll K, et al. Progression-free survival is a
  surrogate for survival in advanced colorectal cancer. {\it Journal of
  Clinical Oncology} 2007\string; 25(33)\string: 5218--5224.

\bibitem{giessen2013progression}
Giessen C, Laubender RP, Ankerst DP, et al. Progression-free survival as a
  surrogate endpoint for median overall survival in metastatic colorectal
  cancer: literature-based analysis from 50 randomized first-line trials. {\it
  Clinical Cancer Research} 2013\string; 19(1)\string: 225--235.

\bibitem{ciani2015meta}
Ciani O, Buyse M, Garside R, et al. Meta-analyses of randomized controlled
  trials show suboptimal validity of surrogate outcomes for overall survival in
  advanced colorectal cancer. {\it Journal of clinical epidemiology}
  2015\string; 68(7)\string: 833--842.

\bibitem{chirila2012meta}
Chirila C, Odom D, Devercelli G, et al. Meta-analysis of the association
  between progression-free survival and overall survival in metastatic
  colorectal cancer. {\it International journal of colorectal disease}
  2012\string; 27(5)\string: 623--634.

\bibitem{KRASmut2017}
Macedo M, M~Melo F, Ribeiro H, et al. KRAS mutation status is highly
  homogeneous between areas of the primary tumor and the corresponding
  metastasis of colorectal adenocarcinomas: One less problem in patient care.
  2017\string; 7\string: 1978-1989.

\bibitem{heterogeneityofCrC2013}
Perez K, Walsh R, Brilliant KE, et al. Heterogeneity of colorectal cancer (CRC)
  in reference to KRAS proto-oncogene utilizing wave technology.. {\it Journal
  of Clinical Oncology} 2013\string; 31(15\_suppl)\string: e14637-e14637.

\bibitem{efron1975}
Efron B, Morris C. Data Analysis Using Stein's Estimator and Its
  Generalizations.  1975\string; 70\string: 311-319.

\bibitem{louis1984}
Louis TA. Estimating a Population of Parameter Values Using Bayes and Empirical
  Bayes Methods. {\it Journal of the American Statistical Association}
  1984\string; 79(386)\string: 393-398.

\bibitem{louis2000}
Guadalupe G. 1. Bayes and Empirical Bayes Methods for Data Analysis. 2nd edn.
  Bradley P. Carlin and Thomas A. Louis. Chapman and Hall/CRC, Hertfordshire,
  U.K., 2000. No. of pages: xvii + 419. Price: 34.99. ISBN 1-58488-170-4. {\it
  Statistics in Medicine}\string; 21(23)\string: 3751-3752.
\newblock \href {\doibase 10.1002/sim.1230} {doi: 10.1002/sim.1230}

\bibitem{neuenschwander2016robust}
Neuenschwander B, Wandel S, Roychoudhury S, Bailey S. Robust exchangeability
  designs for early phase clinical trials with multiple strata. {\it
  Pharmaceutical statistics} 2016\string; 15(2)\string: 123--134.

\bibitem{bland1999measuring}
Bland JM, Altman DG. Measuring agreement in method comparison studies. {\it
  Statistical methods in medical research} 1999\string; 8(2)\string: 135--160.

\bibitem{berry1990SubgroupAnal}
Berry DA. Subgroup Analyses. {\it Biometrics} 1990\string; 46(4)\string:
  1227--1230.

\bibitem{grouin2005Subgroups}
Grouin JM, Coste M, Lewis J. Subgroup Analyses in Randomized Clinical Trials:
  Statistical and Regulatory Issues. {\it Journal of Biopharmaceutical
  Statistics} 2005\string; 15(5)\string: 869-882.
\newblock PMID: 16078390\href {\doibase 10.1081/BIP-200067988} {doi:
  10.1081/BIP-200067988}

\bibitem{kass1995reference}
Kass RE, Wasserman L. A reference Bayesian test for nested hypotheses and its
  relationship to the Schwarz criterion. {\it Journal of the american
  statistical association} 1995\string; 90(431)\string: 928--934.

\bibitem{verdinelli1995computing}
Verdinelli I, Wasserman L. Computing Bayes factors using a generalization of
  the Savage-Dickey density ratio. {\it Journal of the American Statistical
  Association} 1995\string; 90(430)\string: 614--618.

\bibitem{jeffreys1998theory}
Jeffreys H. {\it The theory of probability}.
\newblock OUP Oxford .
\newblock 1998.

\bibitem{berry2013Hierarchical}
Berry SM, Broglio KR, Groshen S, Berry DA. Bayesian hierarchical modeling of
  patient subpopulations: Efficient designs of Phase II oncology clinical
  trials. {\it Clinical Trials} 2013\string; 10(5)\string: 720-734.
\newblock PMID: 23983156\href {\doibase 10.1177/1740774513497539} {doi:
  10.1177/1740774513497539}

\bibitem{thall2003}
F. TP, Kyle WJ, Nebiyou BB, E. CR, H. BL, S. BR. Hierarchical Bayesian
  approaches to phase II trials in diseases with multiple subtypes. {\it
  Statistics in Medicine} 2003\string; 22(5)\string: 763-780.
\newblock \href {\doibase 10.1002/sim.1399} {doi: 10.1002/sim.1399}

\bibitem{chugh2009phase}
Chugh R, Wathen JK, Maki RG, et al. Phase II multicenter trial of imatinib in
  10 histologic subtypes of sarcoma using a bayesian hierarchical statistical
  model. {\it J Clin Oncol} 2009\string; 27(19)\string: 3148--53.

\bibitem{R2w}
Sturtz S, Ligges U, Gelman A. R2WinBUGS: A Package for Running WinBUGS from R.
  {\it Journal of Statistical Software} 2005\string; 12(3)\string: 1--16.

\bibitem{therasse2000Tumors}
Therasse P, Arbuck SG, Eisenhauer EA, et al. New Guidelines to Evaluate the
  Response to Treatment in Solid Tumors. {\it JNCI: Journal of the National
  Cancer Institute} 2000\string; 92(3)\string: 205-216.

\bibitem{world1979handbook}
Organization WH, others . WHO handbook for reporting results of cancer
  treatment.  1979.

\bibitem{bennouna2013continuation}
Bennouna J, Sastre J, Arnold D, et al. Continuation of bevacizumab after first
  progression in metastatic colorectal cancer (ML18147): a randomised phase 3
  trial. {\it The lancet oncology} 2013\string; 14(1)\string: 29--37.

\bibitem{rothenberg2008capecitabine}
Rothenberg M, Cox J, Butts C, et al. Capecitabine plus oxaliplatin (XELOX)
  versus 5-fluorouracil/folinic acid plus oxaliplatin (FOLFOX-4) as second-line
  therapy in metastatic colorectal cancer: a randomized phase III
  noninferiority study. {\it Annals of Oncology} 2008\string; 19(10)\string:
  1720--1726.

\bibitem{cassidy2011xelox}
Cassidy J, Clarke S, D{\'\i}az-Rubio E, et al. XELOX vs FOLFOX-4 as first-line
  therapy for metastatic colorectal cancer: NO16966 updated results. {\it
  British journal of cancer} 2011\string; 105(1)\string: 58.

\bibitem{hurwitz2004bevacizumab}
Hurwitz H, Fehrenbacher L, Novotny W, et al. Bevacizumab plus irinotecan,
  fluorouracil, and leucovorin for metastatic colorectal cancer. {\it New
  England journal of medicine} 2004\string; 350(23)\string: 2335--2342.

\bibitem{mcneish2016modeling}
McNeish D, Stapleton LM. Modeling clustered data with very few clusters. {\it
  Multivariate behavioral research} 2016\string; 51(4)\string: 495--518.

\bibitem{latimer2016treatment}
Latimer NR, Henshall C, Siebert U, Bell H. Treatment switching: statistical and
  decision-making challenges and approaches. {\it International journal of
  technology assessment in health care} 2016\string; 32(3)\string: 160--166.

\bibitem{shanafelt2004chemotherapy}
Shanafelt TD, Loprinzi C, Marks R, Novotny P, Sloan J. Are chemotherapy
  response rates related to treatment-induced survival prolongations in
  patients with advanced cancer. {\it Journal of clinical oncology}
  2004\string; 22(10)\string: 1966--1974.

\bibitem{bujkiewicz2019bivariate}
Bujkiewicz S, Jackson D, Thompson JR, et al. Bivariate network meta-analysis
  for surrogate endpoint evaluation. {\it Statistics in medicine} 2019\string;
  38(18)\string: 3322--3341.

\bibitem{alonso2016applied}
Alonso A, Bigirumurame T, Burzykowski T, et al. {\it Applied surrogate endpoint
  evaluation methods with SAS and R}.
\newblock CRC Press .
\newblock 2016.

\bibitem{bujkiewicz2015bayesian}
Bujkiewicz S, Thompson JR, Riley RD, Abrams KR. Bayesian meta-analytical
  methods to incorporate multiple surrogate endpoints in drug development
  process. {\it Statistics in medicine} 2015\string; 35(7)\string: 1063--1089.

\bibitem{kaizer2017bayesian}
Kaizer AM, Koopmeiners JS, Hobbs BP. Bayesian hierarchical modeling based on
  multisource exchangeability. {\it Biostatistics} 2017\string; 19(2)\string:
  169--184.

\end{thebibliography}

\end{document}